\documentclass{iopart}
\usepackage{graphicx}
\usepackage{amssymb}

\newcommand{\wb}{\omega_{\mathrm{b}}}
\newcommand{\wn}{\omega_{\mathrm{n}}}
\newcommand{\wm}{\omega_{\mathrm{m}}}
\newcommand{\wo}{\omega_{\mathrm{o}}}
\newcommand{\wno}{\omega_{\mathrm{n0}}}
\newcommand{\no}{n_0}
\newcommand{\noo}{ {n{\small 0}}}
\newcommand{\wt}{\wb\, t}

\newcommand{\ee}{\varepsilon}

\newcommand{\sign}{\mathrm{sign}}

\newcommand{\fracc}[2]{\frac{\displaystyle #1}{\displaystyle #2}}

\newcommand{\middlefig}{.5\textwidth}

\newcommand{\s}{\sigma}
\newcommand{\h}{\eta}
\newcommand{\g}{\gamma}
\newcommand{\p}{\varphi}
\newcommand{\la}{\lambda}
\renewcommand{\k}{\kappa}
\renewcommand{\d}{\delta}
\newcommand{\cn}{\mathrm{cn}}
\newcommand{\cd}{\mathrm{cd}}

\newcommand{\In}{\mathrm{In}}

\newcommand{\dd}{\mathrm{d}}

\begin{document}

\title[Influence of impurities on the stability of multibreathers]
{Effect of the introduction of impurities on the stability properties
of multibreathers at low coupling}

\author{J Cuevas, JFR Archilla and FR Romero}

\address{Nonlinear Physics Group of the University of Sevilla
\\ Departamento de F\'{\i}sica Aplicada I, ETSI Inform\'atica\\
Avda Reina Mercedes s/n, 41012 Sevilla, Spain}

\begin{abstract}
Using a theorem dubbed the {\em Multibreather Stabiliy Theorem}
[Physica D 180 (2003) 235-255] we have obtained the stability
properties of multibreathers in systems of coupled oscillators
with on-site potentials, with an inhomogeneity. Analytical results
are obtained for 2-site, 3-site breathers, multibreathers,
phonobreathers and dark breathers. The inhomogeneity is considered
both at the on-site potential and at the coupling terms. All the
results have been checked numerically with excellent agreement.
The main conclusion is that the introduction of a impurity does
not alter the stability properties.

\end{abstract}

\ead{jcuevas@us.es}

\submitto{\NL}

\section{Introduction}
\label{sec:introduction}

Discrete breathers (DBs) are periodic, localized solutions that
appear in discrete lattices of nonlinear oscillators. For
Klein-Gordon lattices, i.e, lattices with an on-site potential,
the conditions for their existence at low coupling are very weak.
They are based on properties of the lattice at the {\em anti
continuous} limit, that is, the same lattice with zero coupling
and, then, with the oscillators isolated. These conditions can be
expressed in simple words as: a)~For a given frequency $\wb$ the
isolated oscillator is truly nonlinear ($\dd E/\dd\wb\neq 0$, with
$E$ the energy of the isolated oscillator). b)~The breather does
not resonate with the phonons ($n\,\wb\neq \wo$, with $n$ any
positive integer and $\wo$ the frequency of the isolated
oscillators~\cite{MA94}. The stability of one--site breathers has
been proven in~\cite{A97, MAF98,MS98}, whereas the stability of
two-sites breathers was proven in Ref.~\cite{MARIN}. Recently,
some of the authors developed the {\em Multibreathers Stability
Theorem} (MST)~\cite{ACSA03}, which provides with a method for
obtaining the stability properties of any multibreather at
low-coupling and applied it to homogeneous lattices. Although the
MST applies to any Klein-Gordon lattice, it depends on the
analytical or numerical calculation of some magnitudes $J_{nm}$ :
This calculation becomes less simple and the number of magnitudes
larger as the system becomes more complicated, involving, for
example, different linear frequencies or initial phases for the
oscillators.

In this paper we address the application of the MST to
Klein-Gordon lattices with an impurity. This impurity can be
modelled at the on-site potential or at the coupling. The
implementation of it at the masses is equivalent to at the on-site
potential. The necessary magnitudes are calculated and the MST
applied to 2-sites and 3-site breathers, multibreathers,
phonobreathers and dark breathers. The theoretical results are
also checked numerically to confirm the validity of the
calculations with excellent results.

The paper is organized in the following form: first we describe
the details of the model; second we recall the Multibreathers
Stability Theorem, and give some details of its application to the
case of a lattice with an impurity. Subsequent sections
particularize it for two and three--sites breathers,
phonobreathers, multibreathers and dark breathers. Details of
analytical calculations are relegated to the first appendix. In
the second appendix, a conjecture made in Ref.~\cite{ACSA03} is
proven. It is that the only multibreathers that can be stable are
the ones with all the oscillator in phase or out of phase.

\section{The model}
\label{sec:model}

In this paper, we consider Klein--Gordon chains with linear
nearest-neighbours coupling whose dynamical equations are of the
form:

\begin{equation}
\ddot{u}_n \,+\,V'(u_n)\,+\,\ee \,\sum_{m=1}^N
\,C_{nm}u_n\,=\,0\,\quad n=1,\dots,N \label{eq:klein}
\end{equation}
where the variables $u_n$ are the displacements with respect to
the equilibrium positions, $V(u_n)$ is the on--site potential, $N$
is the number of oscillators, $C$ is a coupling matrix which
includes the boundary conditions, and $\ee$ is the coupling
parameter.

The impurity can be introduced in the system through an
inhomogeneity at the on--site potential or the coupling matrix, in
a similar fashion as it was done in~\cite{CPAR02b}.

The linear frequencies of the isolated oscillators are
$\wn=(V_n''(0))^{1/2}$. If the inhomogeneity is implemented at the
potential of the $\no$-th particle, we denote $V_0=V_n$ and  $\wo
=\wn=(V_0''(0))^{1/2}$ $\forall n\neq\no$. We can write
$\wno^2=(1+\alpha\delta_{n,\no})\,\wo$, where the inhomogeneity
parameter $\alpha$ takes values in $(-1,\infty)$, $\alpha=0$
corresponding to the homogeneous case.

For a system with nearest-neighbour interaction, the dynamical
equations Eq.~(\ref{eq:klein}) can be written \cite{CPAR02b}:
\begin{equation}
\ddot{u}_n
\,+\,V'(u_n)\,+\,\ee\,[\,k_{n-1,n}\,(u_n-u_{n-1})+k_{n,n+1}\,(u_n-u_{n+1})\,]
=\,0\,. \label{eq:coupling}
\end{equation}
If the system is homogeneous $k_{n,m}=1$, $\forall n,m$, except at
$m=1$ or $n=N$ for a system with $N$ oscillators. With periodic
boundary conditions, the index $n$ is cyclical so as $0\sim N$ and
$N+1\sim 1$ and $k_{0,1}=k_{N,N+1}=1$. With free ends,
$k_{0,1}=k_{N,N+1}=0$. With fixed ends, two extra oscillators have
$u_0=u_{N+1}=0$ and $k_{0,1}=k_{N,N+1}=1$.

Therefore, the coupling matrix, $C^0$ for a finite system with $N$
oscillators with fixed ends (at the oscillators $n=0$ and
$n=N+1$), has elements:
\begin{equation}
C^0_{nm}=\left\{\begin{array}{ll} 2 & \textrm{if } n=m \\ -1 &
\textrm{if } |n-m|=1 \\0 & \textrm{otherwise}.\end{array}\right.
\end{equation}
With  free ends, $C_{1\,1}=C_{N\,N}=1$, and with  periodic
boundary conditions, $C_{1\,N}=C_{N\,1}=-1$.

We model the impurity at a site $n_0$ distinct from the lattice
boundaries, by changing the constants $k_{\noo,\noo+1}$ and
$k_{\noo-1,\noo}$ to $(1+\beta/2)$, with $\beta$ being a parameter
which takes its values in $(-2,\infty)$, so as the homogeneous
case is recovered when $\beta=0$. $\beta<0$ means weaker constants
than the homogeneous coupling ones and $\beta>0$ the opposite.

Therefore, the coupling matrix $C_{nm}$ becomes:
\begin{equation}
C_{nm}=\left\{\begin{array}{ll}
(\beta+2) & \textrm{if } n=m=\no \\
-(\beta+2)/2 & \textrm{if } |n-m|=1 \textrm{ and } n=\no \textrm{
or }
m=\no \\
(\beta+4)/2 & \textrm{if } n=m=\no\pm1\\
C^0_{nm} & \textrm{otherwise},
\end{array}\right.
\end{equation}

\section{Stability theorem}\label{sec:stab}

In this section we recall some results of the multibreather
stability theorems established in Ref.~\cite{ACSA03}. For more
details about this theory, the reader is referred to that
reference.

The stability of a multibreather at low coupling can be
established through three different parameters: the sign of the
coupling constant $\ee$, the softness/hardness of the on-site
potential and the sign of the eigenvalues $\{\lambda_i\}$ of the
perturbation matrix $Q$, to be described below. Considering a
positive coupling constant, the multibreather is stable for a soft
potential if all the eigenvalues of $Q$ are non-negative ($Q$ is
positive semi-definite); if the potential is soft, the
multibreather is stable if all the eigenvalues $\{\lambda_i\}$ are
negative except for a zero one ($Q$ is negative semi-definite). If
the coupling constant is negative, the results are reversed.
Furthermore, if $Q$ is non-definite (i.e. there are some positive
and some negative eigenvalues), the multibreather is unstable
independently on the sign of $\ee$ and the softness/hardness of
the potential.

We can summarize the stability properties in the following way.
Let $S=1$ mean stability and $S=-1$ instability, $H=1$ correspond
to a hard on-site potential, and $H=-1$ to a soft one, and define
the $\sign(Q)=1$ if all the eigenvalues of $Q$ but a zero one are
positive and $\sign(Q)=-1$ if they are negative except for the
zero one, other cases being indefinite. Then, $S=-H\times
\sign(Q)\times \sign(\ee)$.

It is also important to take into account that there is always a
zero eigenvalue due to a global phase mode. If there were more
than one zero eigenvalue, the stability theorem can only predict
the instability in the case that there existed eigenvalues of
different sign, but not the stability, as the 0--eigenvalue is
degenerate.

The perturbation matrix $Q$ is obtained in the following way:
suppose that at the anticontinuous limit there are $p$ excited
oscillators and $N-p$ oscillators at rest. Let us construct the
modified coupling matrix $\tilde C$ by suppressing in $C$ the rows
and columns corresponding to the $N-p$ oscillators at rest, and
redefine $C=\tilde C$. The dimension of $C$ is $p\times p$, its
nondiagonal elements being \cite{ACSA03}:
\begin{equation}\label{eq:qnm}
    Q_{nm}=\frac{{C}_{nm}}{\mu_n\, \mu_m}\int_{-T/2}^{T/2}\,\dot{u}_n
    \,\dot{u}_m\,\dd t\,, \quad n\neq m\,\,
\end{equation}
with $T=2\,\pi/\wb$. The diagonal elements are:
\begin{equation}\label{eq:qnn}
    Q_{nn}=-\sum_{\forall\, m\neq n}\,\frac{\mu_m}{\mu_n}\,Q_{nm} \,.
\end{equation}
where $u_n$ are the solutions of an isolated oscillator submitted
to the potential $V(u_n)$, i.e. the solutions of the equations:
\begin{equation}
    \ddot u_n+V'(u_n)=0,
    \label{eq:isolated}
\end{equation}
and $\mu_n$ is defined as:
\begin{equation}
    \mu_n=\sqrt{\int_{-T/2}^{T/2}\,\dot{u}^2_n\,\dd t}
\end{equation}

In this paper we limit ourselves to time--reversible solutions,
i.e., the excited oscillators at the anticontinuous limit can only
have phase $0$ or $\pi$, or, in other words, if $u_n(t)$ is a
time--reversible solution of Eq.~(\ref{eq:isolated}), i.e.,
$u_n(t)=u_n(-t)$, then $u_n(t+T/2)$ is the only possible
time--reversible solution apart from $u_n(t)$. Therefore, the
state of the system at the anticontinuous limit can be described
by a code $\sigma=(\sigma_1,\dots\sigma_N)$, where
$\sigma_n=0,+1,-1$ means that the corresponding oscillator is at
rest, with phase 0, or with phase $\pi$, respectively, at
$t=0$~\cite{A97}.

Let $J_{nm}$ and $J'_{nm}$ be defined as:
\begin{equation}
 \fl   J_{nm}=\int_{-T/2}^{T/2}\,\dot{u}_n(t)\,\dot{u}_m(t)\,\dd t\,,
    \quad
    J'_{nm}=\int_{-T/2}^{T/2}\,\dot{u}_n(t)\,\dot{u}_m(t+T/2)\,\dd
    t\,,\quad
\end{equation}
and the parameters $\h_{nm}$, $\g_{nm}$ and $\p_{nm}$ as:
\begin{equation}
\fl \h_{nm}=\fracc{J_{nm}}{\sqrt{J_{nn}J_{mm}}}>0, \quad
    \g_{nm}=-\fracc{J'_{nm}}{\sqrt{J_{nn}J_{mm}}}>0, \quad
    \p_{nm}=\sqrt{\fracc{J_{mm}}{J_{nn}}}>0,\quad \nonumber \label{eq:hgp}
\end{equation}
Then, the matrix elements $Q_{nm}$ can be written as:
\begin{equation}\label{eq:qhgp}
    Q_{nm}=\left\{\begin{array}{ll}\h_{nm}C_{nm} & \textrm{if }
    \s_n\s_m=1 \\ -\g_{nm}C_{nm} & \textrm{if }
    \s_n\s_m=-1\end{array}\right.
\end{equation}
\begin{equation}\label{eq:qhgpd}
     Q_{nn}=-\sum_{\forall\, m\neq n}\p_{nm}Q_{nm},
\end{equation}
where we have taken into account that $\mu_n^2=J_{nn}$. We have
calculated analytically these parameters for several potentials in
\ref{ap:a}, although only the case of the Morse on--site potential
leads to relatively simple expressions.

Some properties of the parameters $\h_{nm}$, $\g_{nm}$ and
$\p_{nm}$ can be easily deduced. In the case of symmetric
potentials, $J_{nm}=-J'_{nm}$, as $\dot u_n(t)=-\dot u_n(t+T/2)$,
and, in consequence, $\g_{nm}=\h_{nm}$. Other property is that,
for the homogeneous case, $\h_{nm}=1$, $\p_{nm}=1$ and
$\g_{nm}=\g_0$, where $\g_0$ is the symmetry parameter in that
case (which is the unity for a symmetric potential). 

In order to simplify the notation, as we are considering a single
impurity, we define the parameters $\h\equiv\h_{nh}$,
$\g\equiv\g_{nh}$, $\p\equiv\p_{nh}$, where the index $n$
indicates the impurity site and $h$ another one.

\section{2-site breathers with an impurity}\label{sec:2site}

A 2-site breather is a breather obtained from the anticontinuous
limit when two contiguous sites are excited. The boundary
conditions are irrelevant as long as the sites are not close to
the boundaries. Without loss of generality, let us rename the two
indexes so as the impurity is located at $n=1$, and the
corresponding oscillator has phase zero ($\sigma_1=1$). The other
oscillator is located at $n=2$. When dealing with the code
$\sigma$ it is enough to consider the codes of these two
oscillators, as the other are zero, thus,
$\sigma=(\sigma_1,\sigma_2)$. Two cases are possible, in--phase
oscillators, $\sigma=(1,1)$, and out--of--phase oscillators,
$\sigma=(1,-1)$.

\subsection{Inhomogeneity at the on-site potential}

First of all, we consider the in-phase case. Taking into account
Eqs. (\ref{eq:hgp}-\ref{eq:qhgpd}) and the notation introduced in
Section \ref{sec:stab}, the matrix $Q$ is given by:
\begin{equation}
    Q=\left[\begin{array}{cc}
    \h\p & -\h \\
    -\h & \h/\p
    \end{array}\right],
\end{equation}
with eigenvalues $\{\la_0,\la_1\}=\{0,\h(\p+1/\p)\}$. Then,
$\la_1>0$.

For  the out-of-phase case, the perturbation matrix is:
\begin{equation}
    Q=\left[\begin{array}{cc}
    -\g\p & \g \\
    \g & -\g/\p
    \end{array}\right],
\end{equation}
with eigenvalues $\{\la_0,\la_1\}=\{0,-\g(\p+1/\p)\}$. Thus,
$\la_1<0$.

If the system is homogeneous  $\{\la_0,\la_1\}=\{0,2\}$ and
$\{\la_0,\la_1\}=\{0,-2\g_0\}$ for the in-phase and out-of-phase
cases, respectively~\cite{ACSA03}. In consequence, the sign of
$\la_1$ does not change when an impurity is introduced in the
on-site potential, and the stability properties are not altered.

\begin{figure}
\begin{center}
\begin{tabular}{cc}
    (a) & (b) \\
    \includegraphics[width=\middlefig]{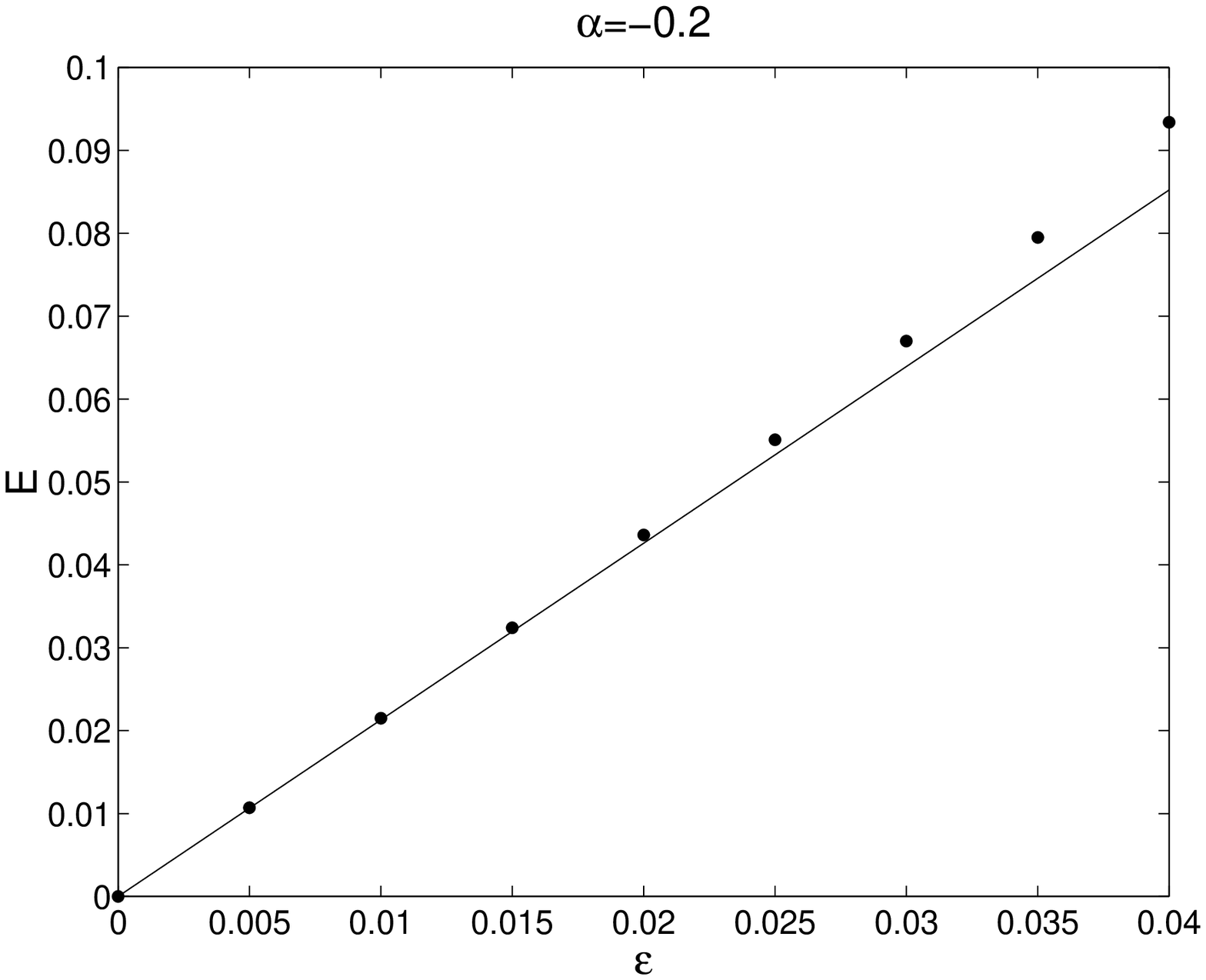} &
    \includegraphics[width=\middlefig]{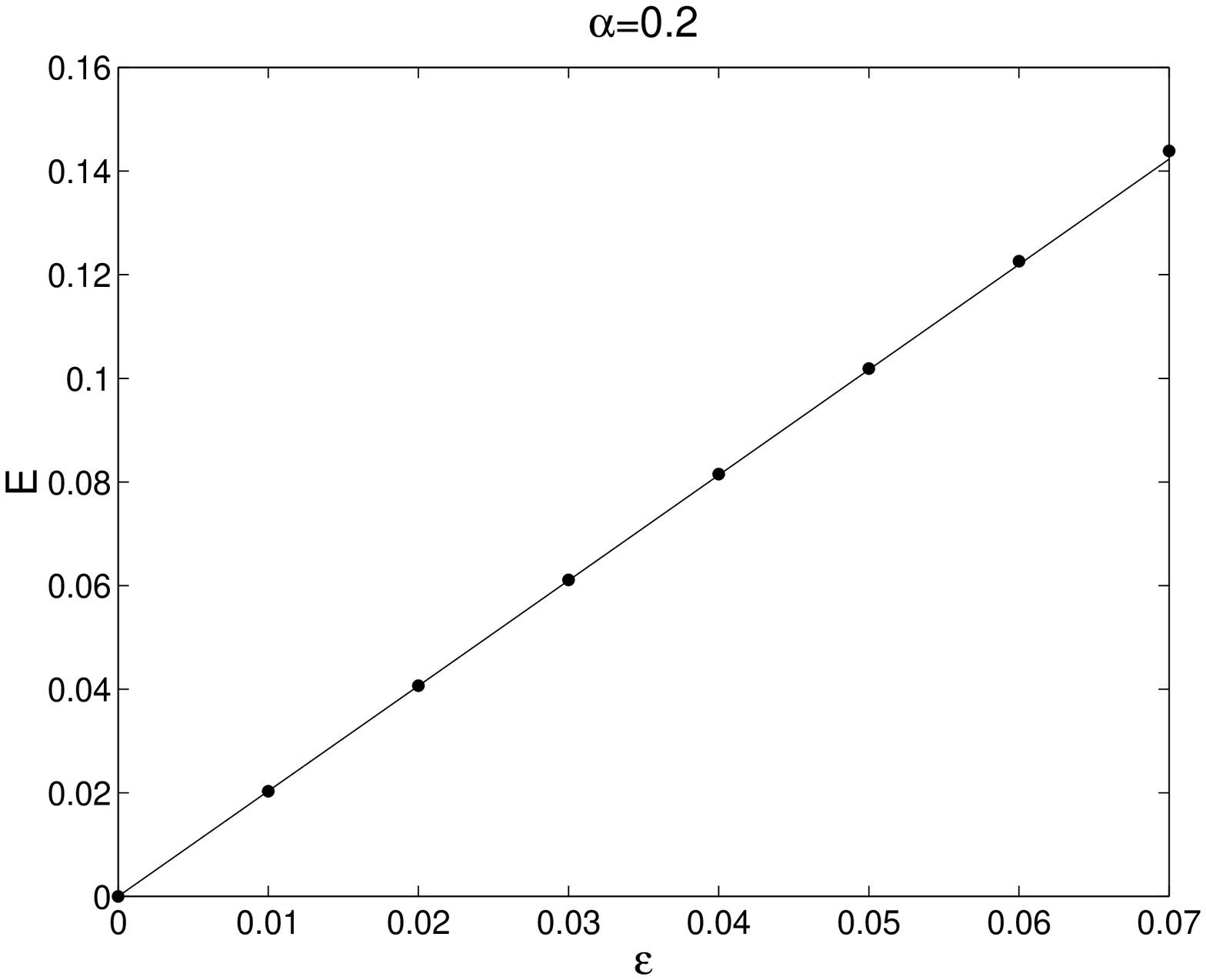}\\
    (c) & (d) \\
    \includegraphics[width=\middlefig]{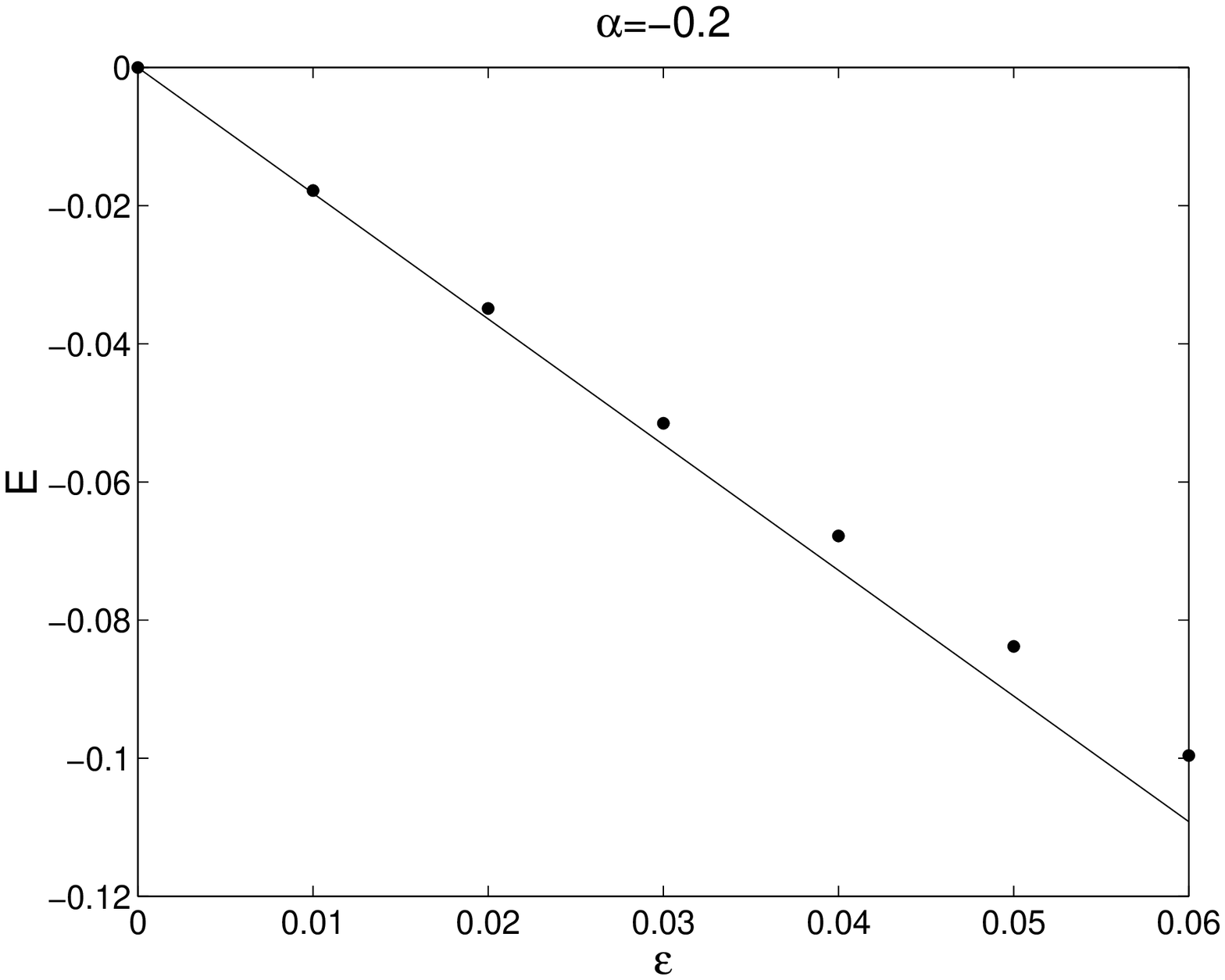} &
    \includegraphics[width=\middlefig]{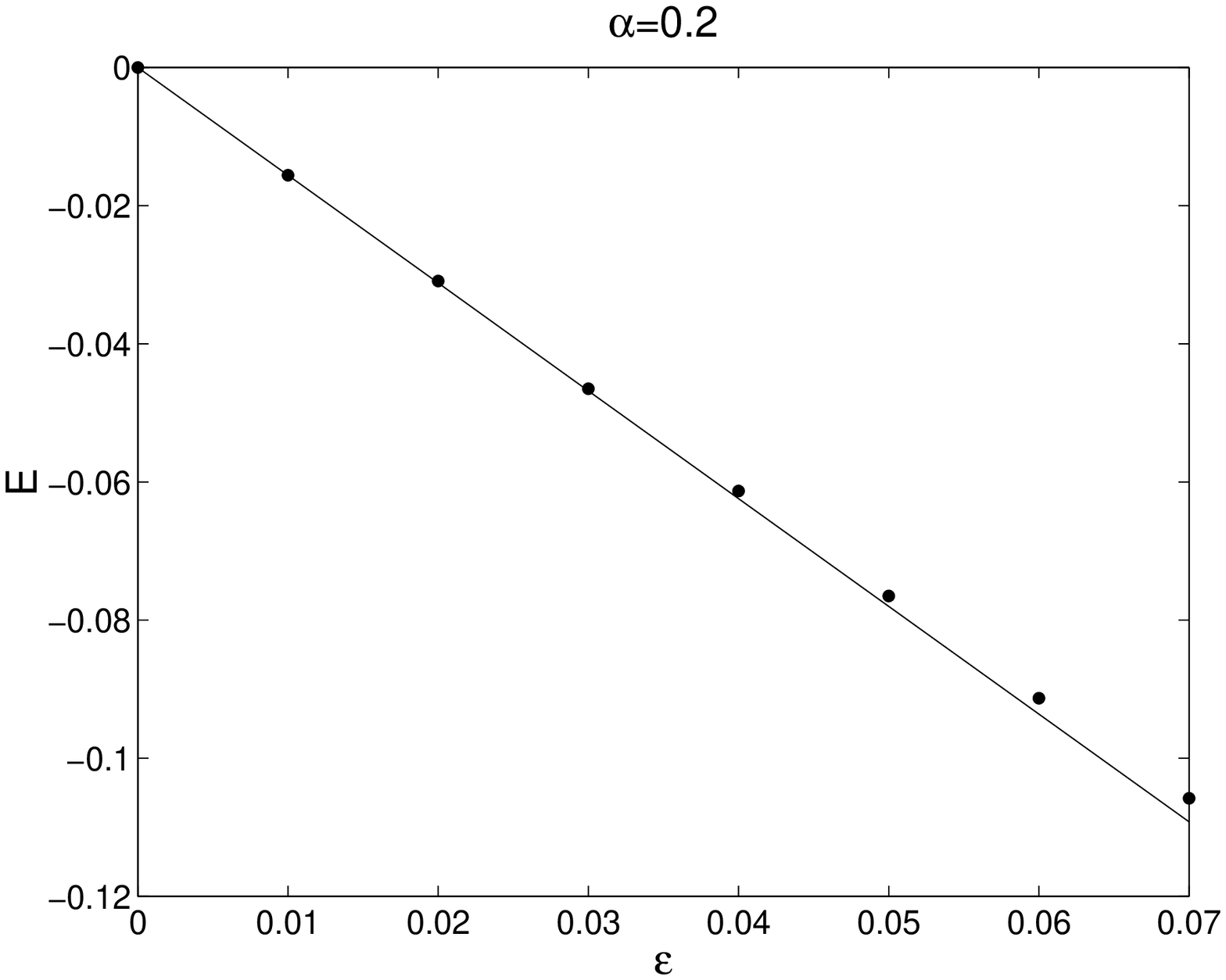}\\
\end{tabular}
\caption{Dependence of $E=\ee\lambda_1$ with respect to $\ee$ for
two different values of $\alpha$. (a,b) correspond to in-phase
2-site breathers and (c,d) to out-of-phase 2-site breathers. All
the solutions stand for a Morse potential and $\wb=0.8$. Dots
represent numerical solutions and lines analytical ones.}%
\label{fig:2sm}
\end{center}
\end{figure}

\begin{figure}
\begin{center}
\begin{tabular}{cc}
    (a) & (b) \\
    \includegraphics[width=\middlefig]{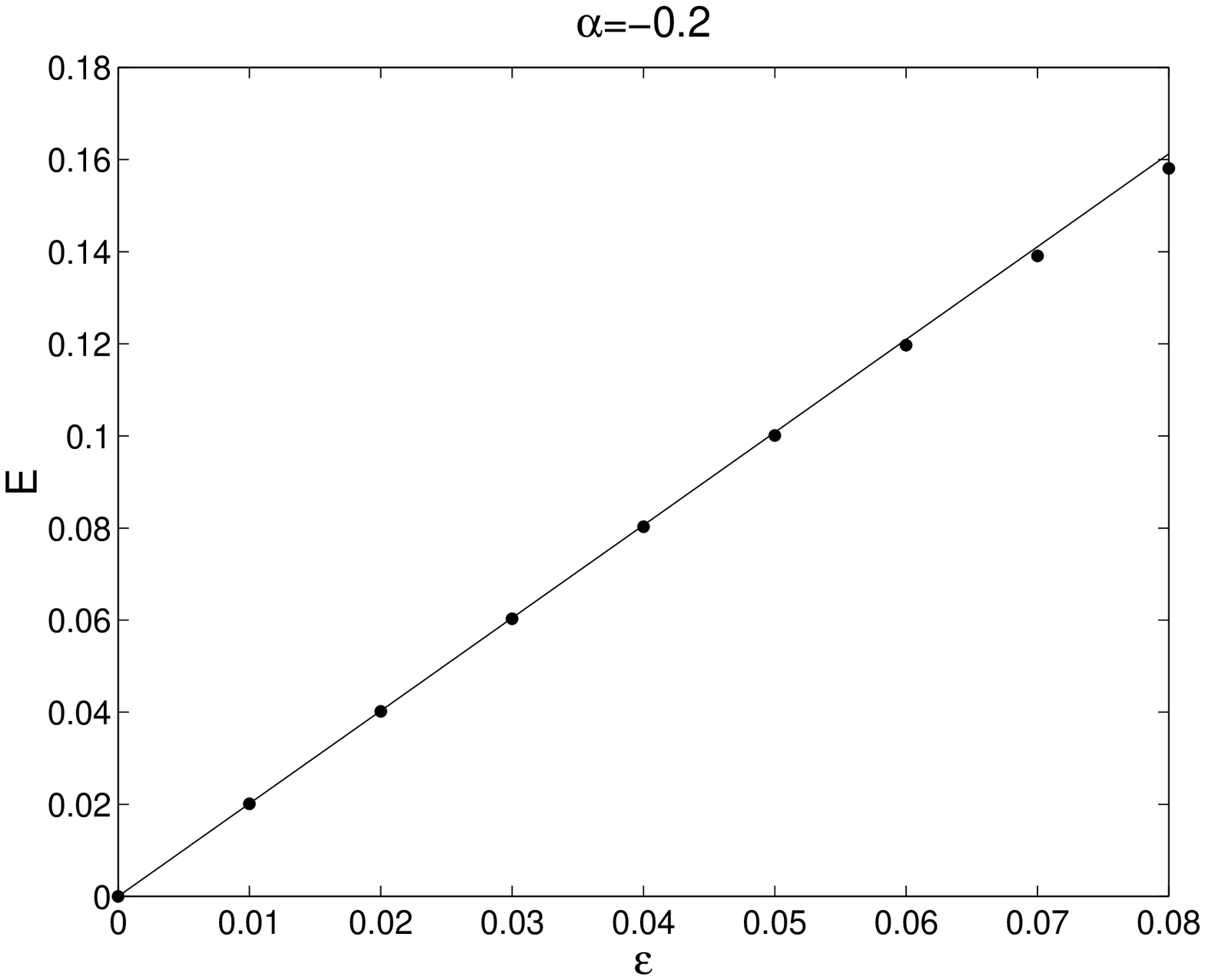} &
    \includegraphics[width=\middlefig]{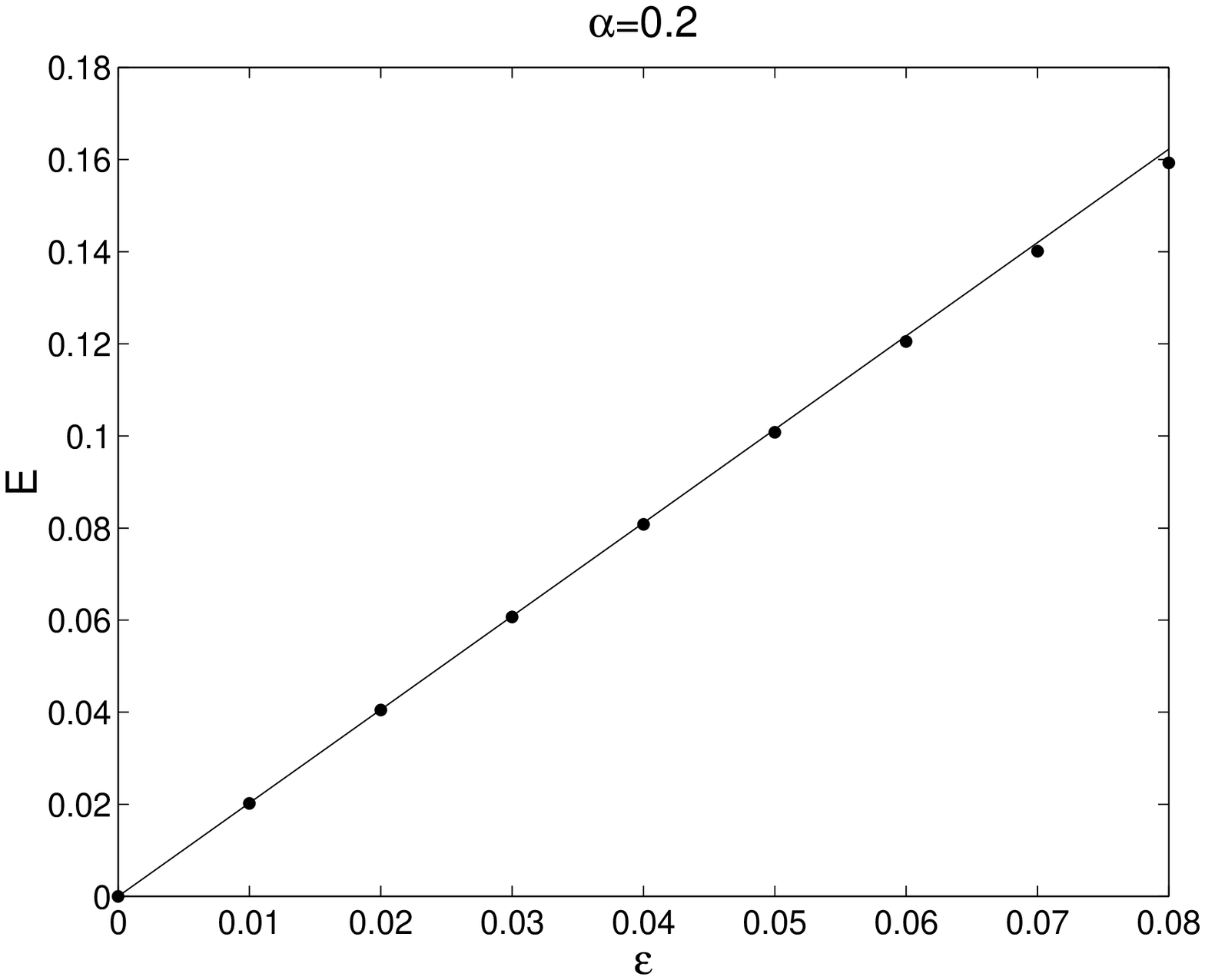}\\
    (c) & (d) \\
    \includegraphics[width=\middlefig]{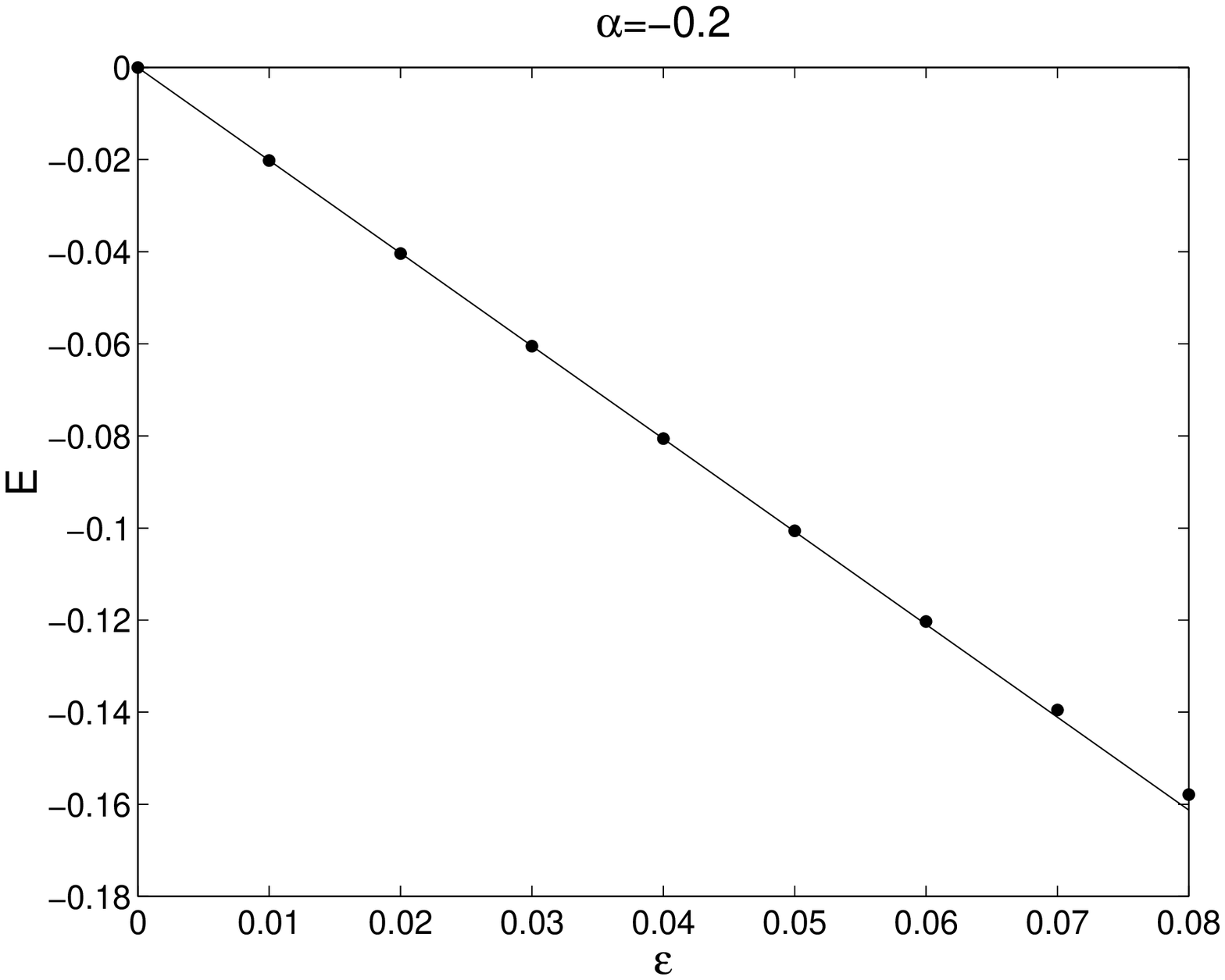} &
    \includegraphics[width=\middlefig]{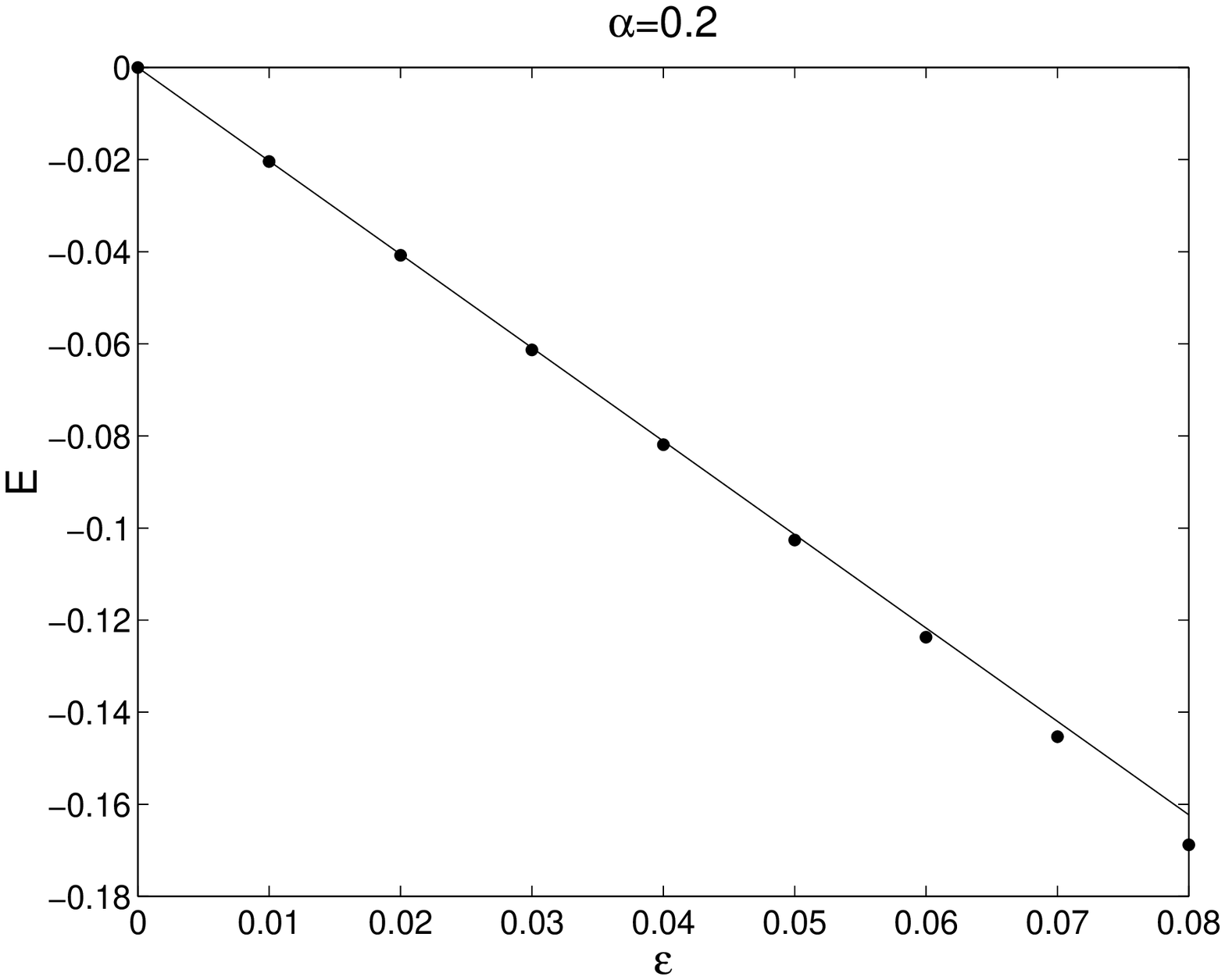}\\
\end{tabular}
\caption{Dependence of $E=\ee\lambda_1$ with respect to $\ee$ for
two different values of $\alpha$. (a,b) correspond to in-phase
2-site breathers and (c,d) to out-of-phase 2-site breathers. All
the solutions stand for a hard $\phi^4$ potential and
$\wb=1.3$. Points represent numerical solutions and lines analytical ones.}%
\label{fig:2sh}
\end{center}
\end{figure}

Figs.~\ref{fig:2sm} and \ref{fig:2sh} show the dependence of
$E=\ee\la_1$ with respect to $\ee$ calculated numerically and
analytically (see \ref{ap:a}). The agreement between both
calculations is excellent for fairly large values of the coupling
constant.

\subsection{Inhomogeneity at the coupling constant}

For the $\sigma=(1,1)$ 2--site breather:
\begin{equation}
    Q=\frac{\beta+2}{2}\left[\begin{array}{cc}
    1 & -1 \\
    -1 & 1
    \end{array}\right]
    =\frac{\beta+2}{2}Q_0,
\end{equation}
where $Q_0$ is the perturbation matrix of the homogeneous chain.
Thus, $\la_1=\beta+2$. The relation $Q=(\beta+2)/2\,Q_0$ also
holds for the code $\sigma=(1,-1)$, and, in consequence,
$\la_1=-(\beta+2)/2\,\g_0$ in this case. As $\beta>-2$, the
conclusion is that the stability properties are the same as in the
homogeneous system.

Fig. \ref{fig:coup} compares this result with the numerically
obtained.

\section{3-site breathers with an impurity}\label{sec:3site}

They consist of breathers derived from three contiguous, excited
oscillators at the anticontinuous limit. There are seven
non--equivalent possibilities, taking into account the position of
the impurity and the phases of the oscillators. Let us rename the
excited sites as $1,2,3$, suppose that $\s_2=1$, and denote with
boldface the position of the impurity. Then the three-index codes
corresponding to the three excited oscillators are (1,{\bf 1},1),
(1,1,{\bf 1}), (-1,{\bf 1},-1), (-1,1,{\bf -1}), ({\bf 1},1,-1),
(1,{\bf 1},-1), (1,1,{\bf -1}).

\subsection{Inhomogeneity at the on-site potential}

Let us suppose that the  $\s\equiv(\s_1,1,\s_3)$. The in-phase
breather corresponds to $\s_1=\s_3=1$, the out-of-phase to
$\s_1=\s_3=-1$ and another one to $\s_1\s_3=-1$. The following
table shows the eigenvalues $\la_1$ and $\la_2$ for the different
configurations ($\la_0=0$ always), the impurity site is indicated
through a bold font, and, in order to simplify the expressions, we
define $\phi=\p+1/\p$:
\begin{center}
\begin{footnotesize}
\begin{tabular}{|c||c|c|}
\hline Code & $\la_1$ & $\la_2$ \\
\hline \hline%
1\ {\bf 1}\ 1 & $\h/\p$ & $\h(\p+\phi)$ \\%
\hline%
1\ 1\ {\bf 1} & $(2+\h\phi+\sqrt{4-4\h\phi+\h^2\phi^2})/2$ &
$(2+\h\phi+\sqrt{4-4\h\phi+\h^2\phi^2})/2$ \\%
\hline%
-1\ {\bf 1}\ -1 & $-\g/\p$ & $-\g(\p+\phi)$ \\%
\hline%
-1\ 1\ {\bf -1} &
$-\g_0-[\g\phi+\sqrt{(\g\phi+2\g_0)^2-4\g\g_0(\p+\phi)}]/2$ &
$-\g_0-[\g\phi-\sqrt{(\g\phi+2\g_0)^2-4\g\g_0(\p+\phi)}]/2$ \\%
\hline%
{\bf 1}\ 1\ -1 & $(\h\phi+\sqrt{\h^2\phi^2+4\g_0\h(\p+\phi)})/2$ &
$(\h\phi-\sqrt{\h^2\phi^2+4\g_0\h(\p+\phi)})/2$ \\%
\hline%
1\ {\bf 1}\ -1 &
$((\h-\g)\phi+\sqrt{(\h+\g)^2\phi^2-4\h\g\p^2})/2$ &
$((\h-\g)\phi-\sqrt{(\h+\g)^2\phi^2-4\h\g\p^2})/2$
\\
\hline%
1\ 1\ {\bf -1} &
$(-(\g\phi-2)+\sqrt{\g^2\phi^2+4+4\g\p})/2$ &
$(-(\g\phi-2)-\sqrt{\g^2\phi^2+4+4\g\p})/2$
\\
\hline
\end{tabular}
\end{footnotesize}
\end{center}
$\g_0$ being the symmetry coefficient for the homogeneous case. It
can be deduced that $\la_1>0$ and $\la_2>0$ for the in-phase
solution, $\la_1<0$ and $\la_2<0$ for the out-of-phase breather
and $\la_1>0$ and $\la_2<0$ otherwise. This result coincides with
the homogeneous case.

\begin{figure}
\begin{center}
\begin{tabular}{cc}
    (a) & (b) \\
    \includegraphics[width=\middlefig]{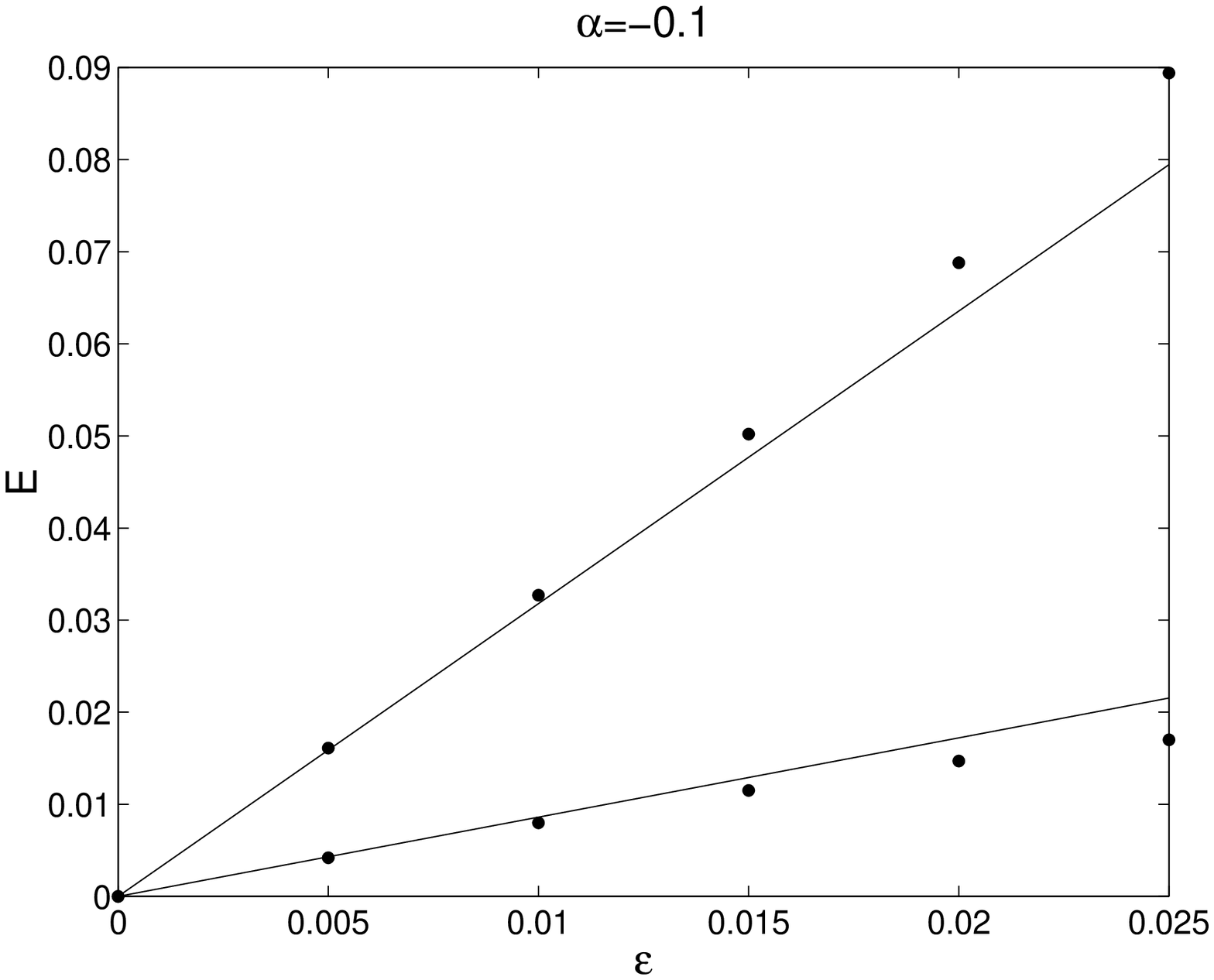} &
    \includegraphics[width=\middlefig]{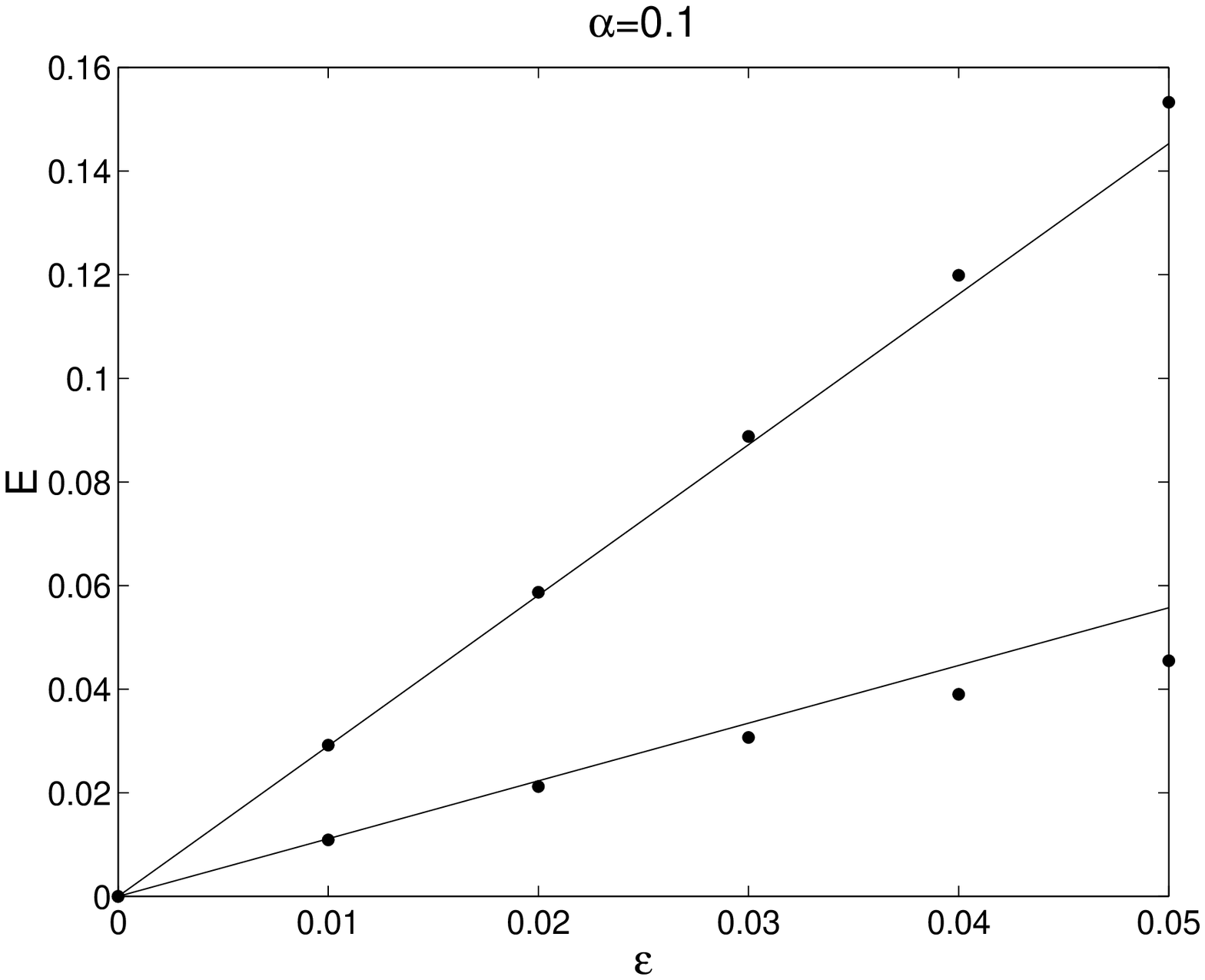}\\
    (c) & (d) \\
    \includegraphics[width=\middlefig]{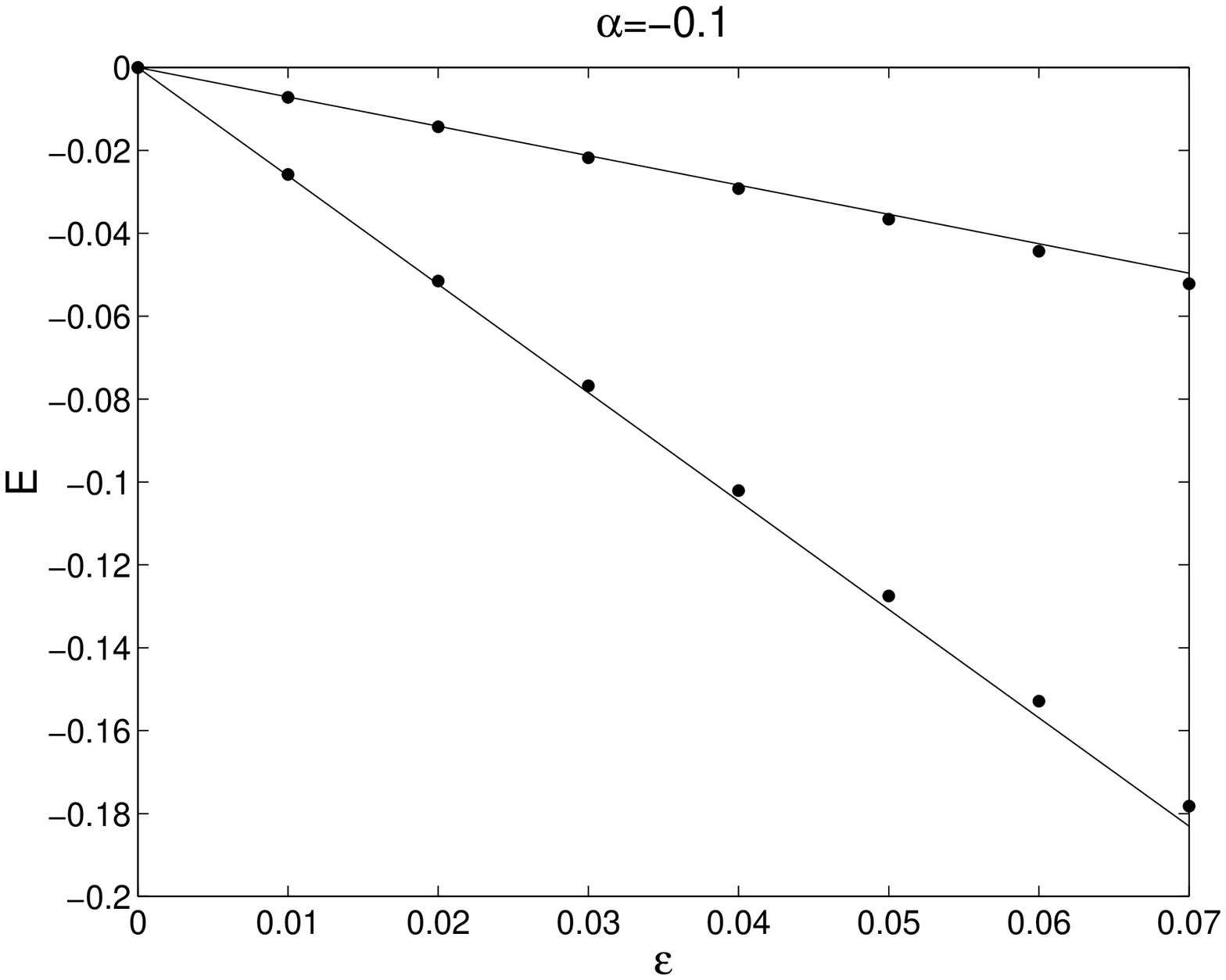} &
    \includegraphics[width=\middlefig]{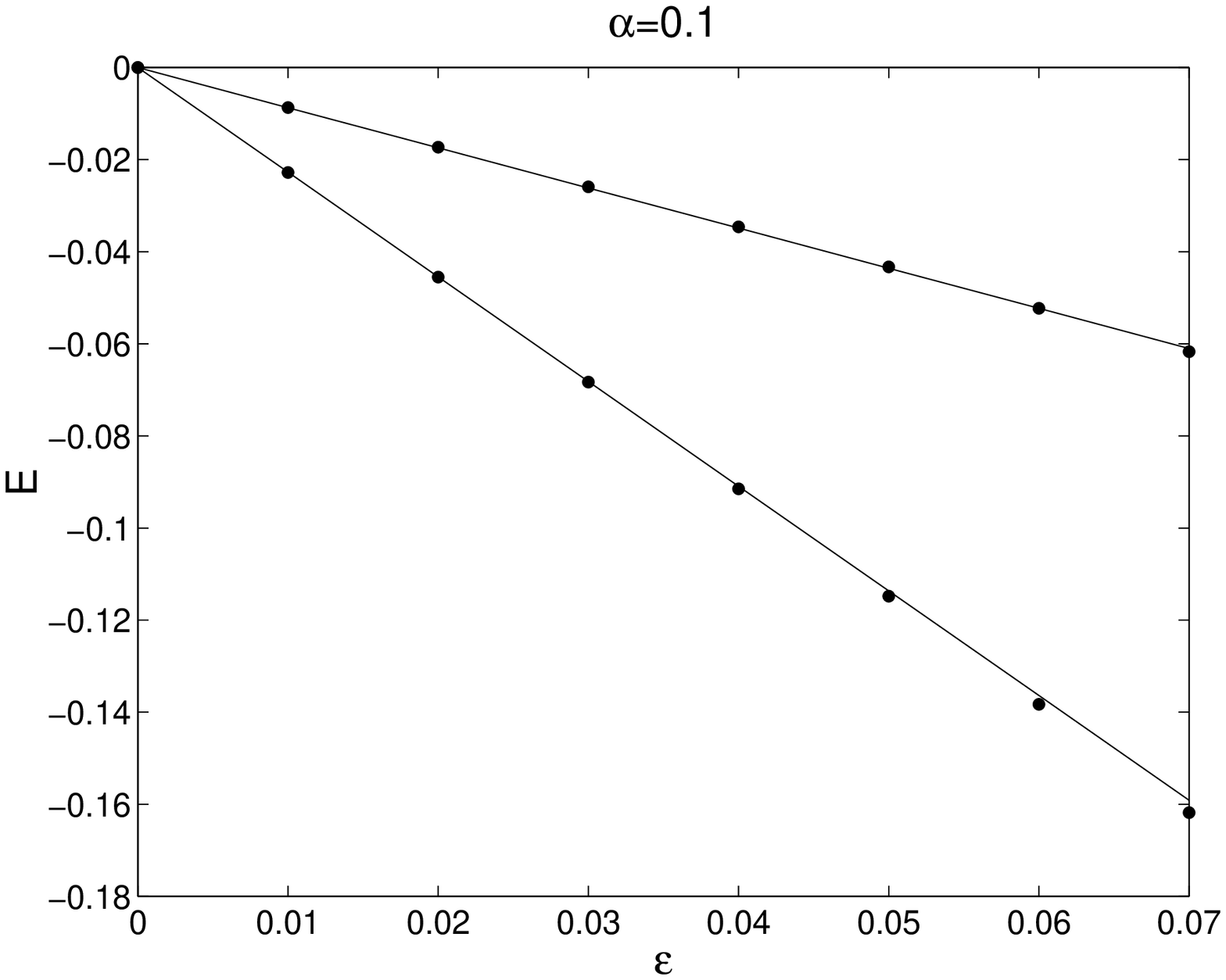}\\
    (e) & (f) \\
    \includegraphics[width=\middlefig]{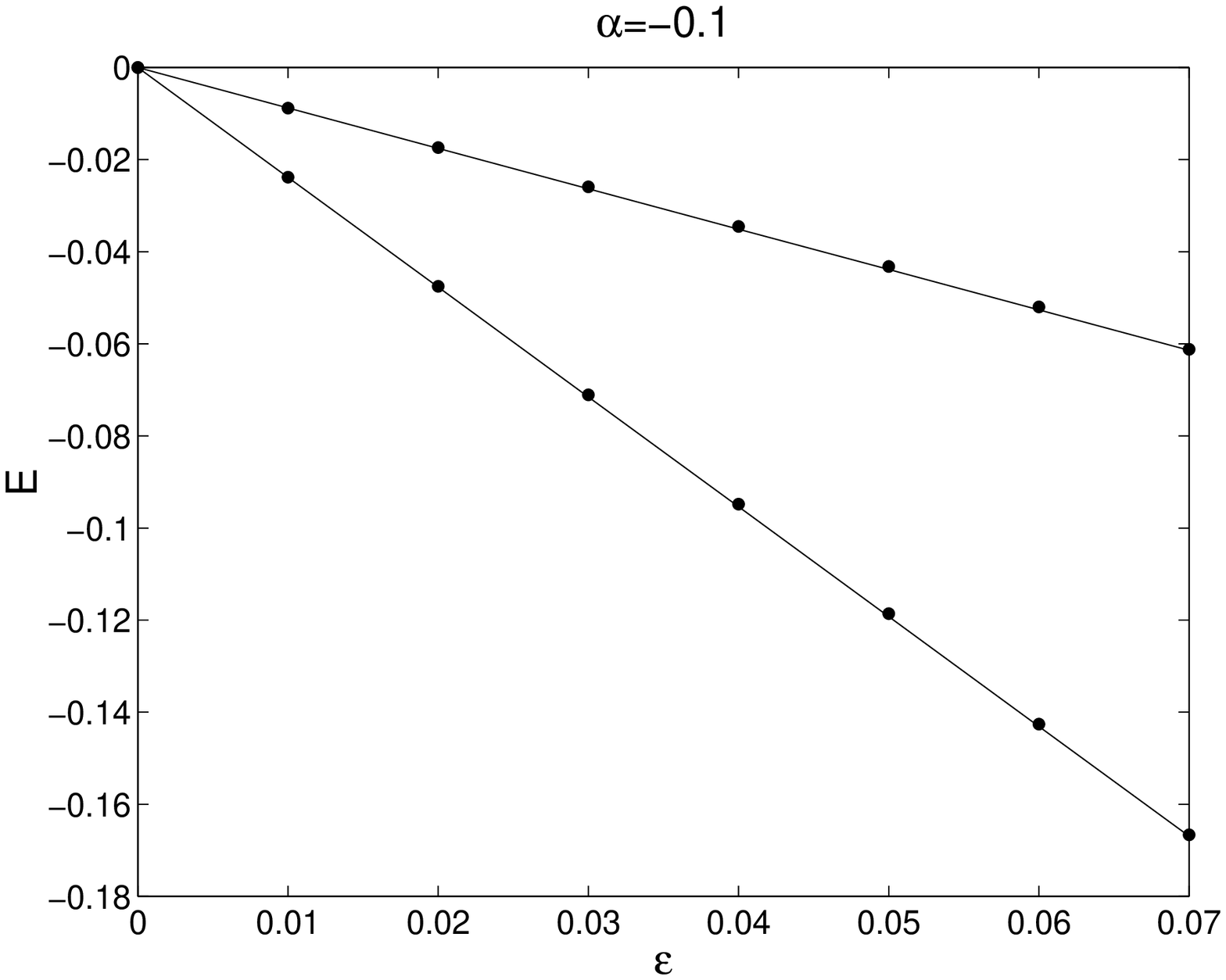} &
    \includegraphics[width=\middlefig]{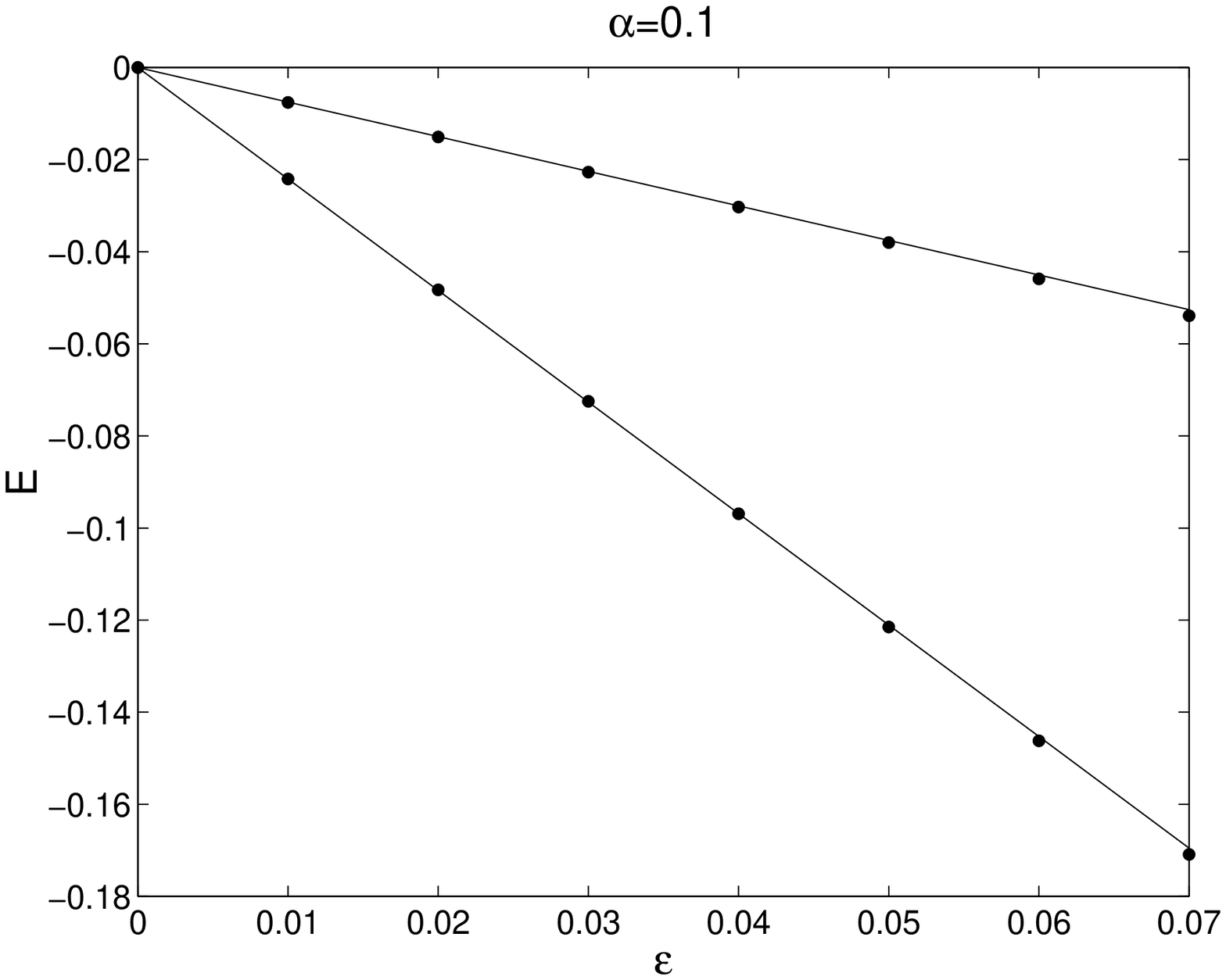}\\
\end{tabular}
\caption{Dependence of $E=\ee\lambda_1$ and $E=\ee\lambda_2$ with
respect to $\ee$ for two different values of $\alpha$. (a,b)
correspond to in-phase 3-site breathers, (c,d) to out-of-phase
3-site breathers with the impurity at the center  and (e,f) to the
same as before but with the impurity at the edge. All the
solutions stand for a Morse potential and $\wb=0.8$. Points
represent numerical solutions and lines analytical ones.}%
\label{fig:3sm}
\end{center}
\end{figure}

\begin{figure}
\begin{center}
\begin{tabular}{cc}
    (a) & (b) \\
    \includegraphics[width=\middlefig]{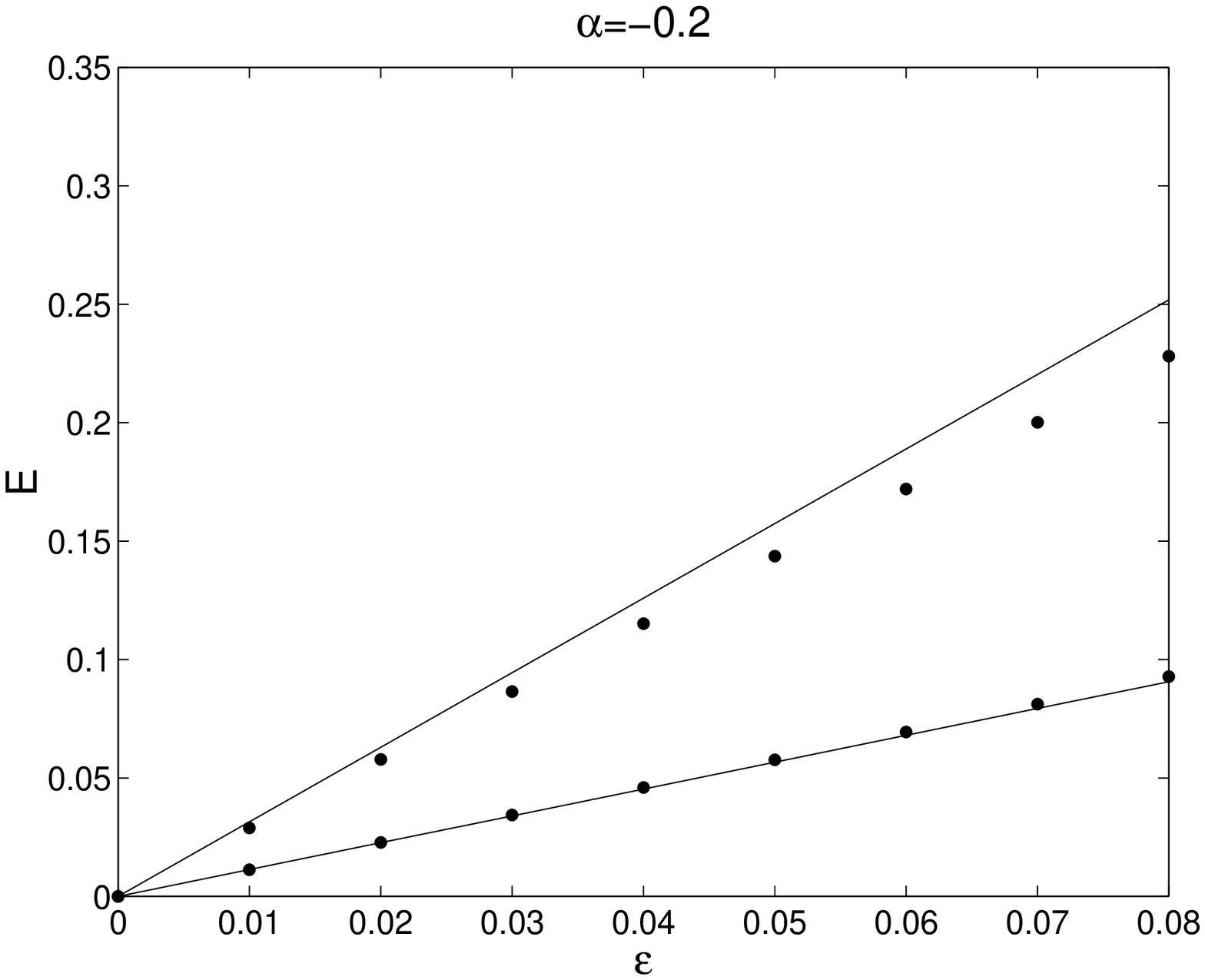} &
    \includegraphics[width=\middlefig]{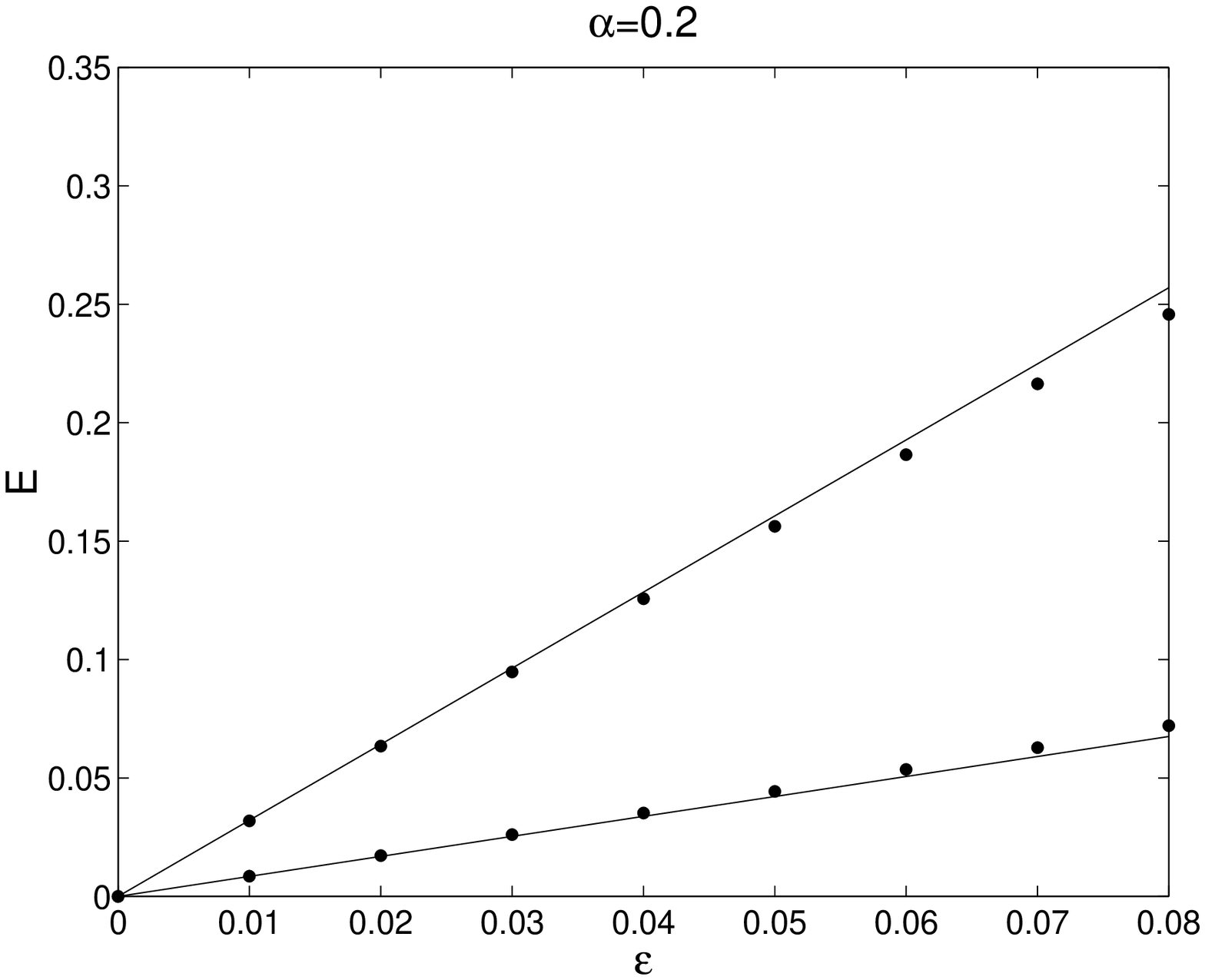}\\
    (c) & (d) \\
    \includegraphics[width=\middlefig]{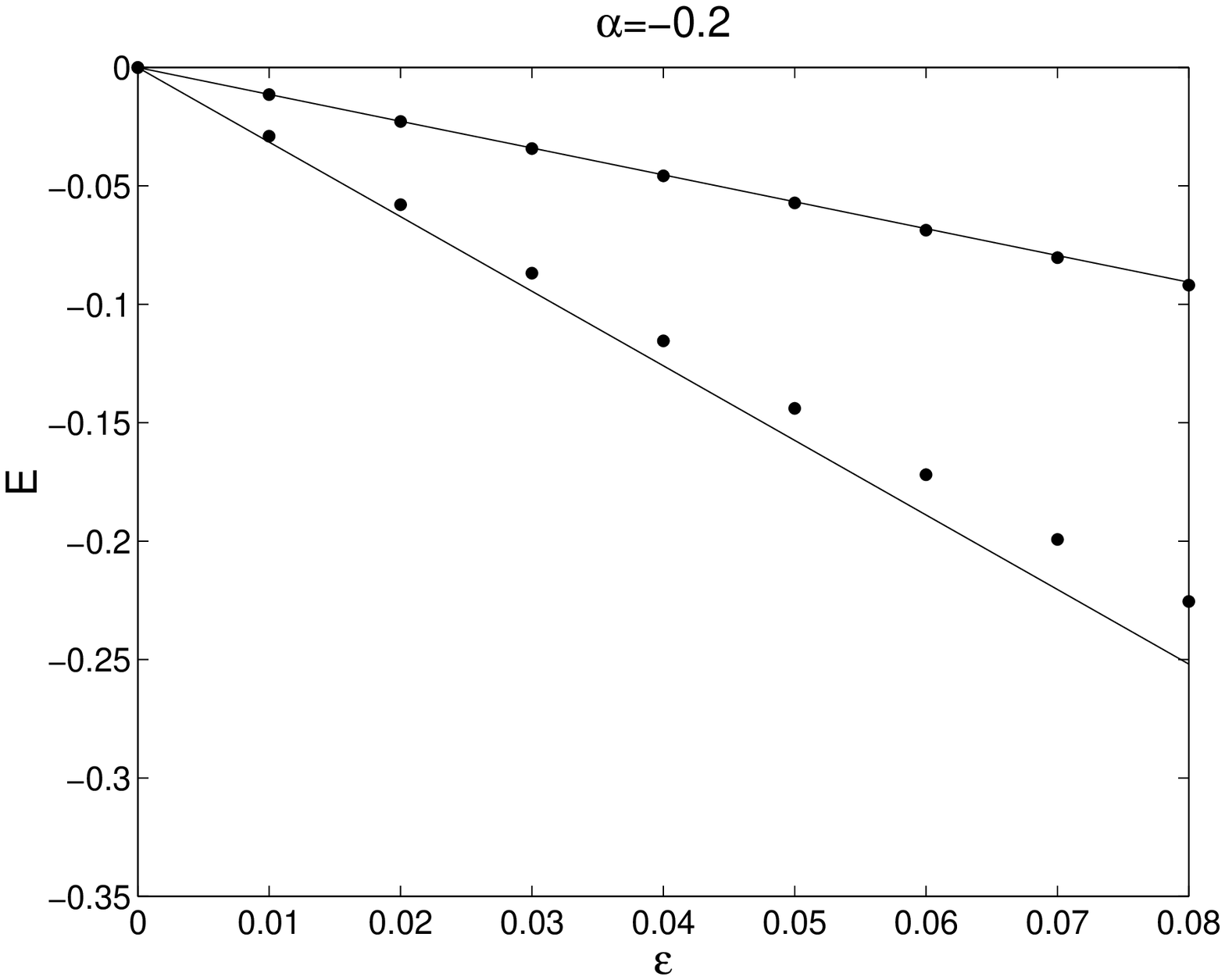} &
    \includegraphics[width=\middlefig]{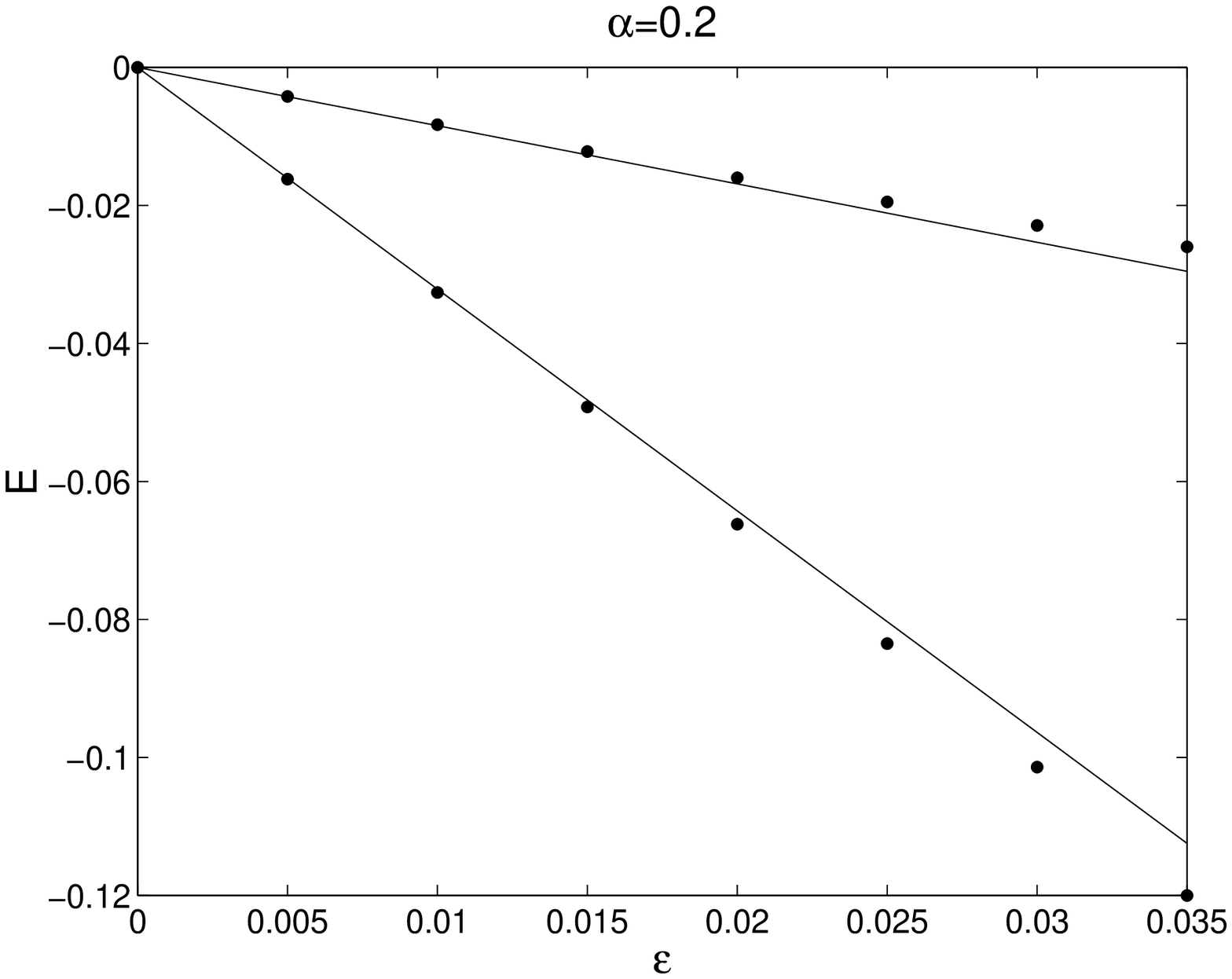}\\
    (e) & (f) \\
    \includegraphics[width=\middlefig]{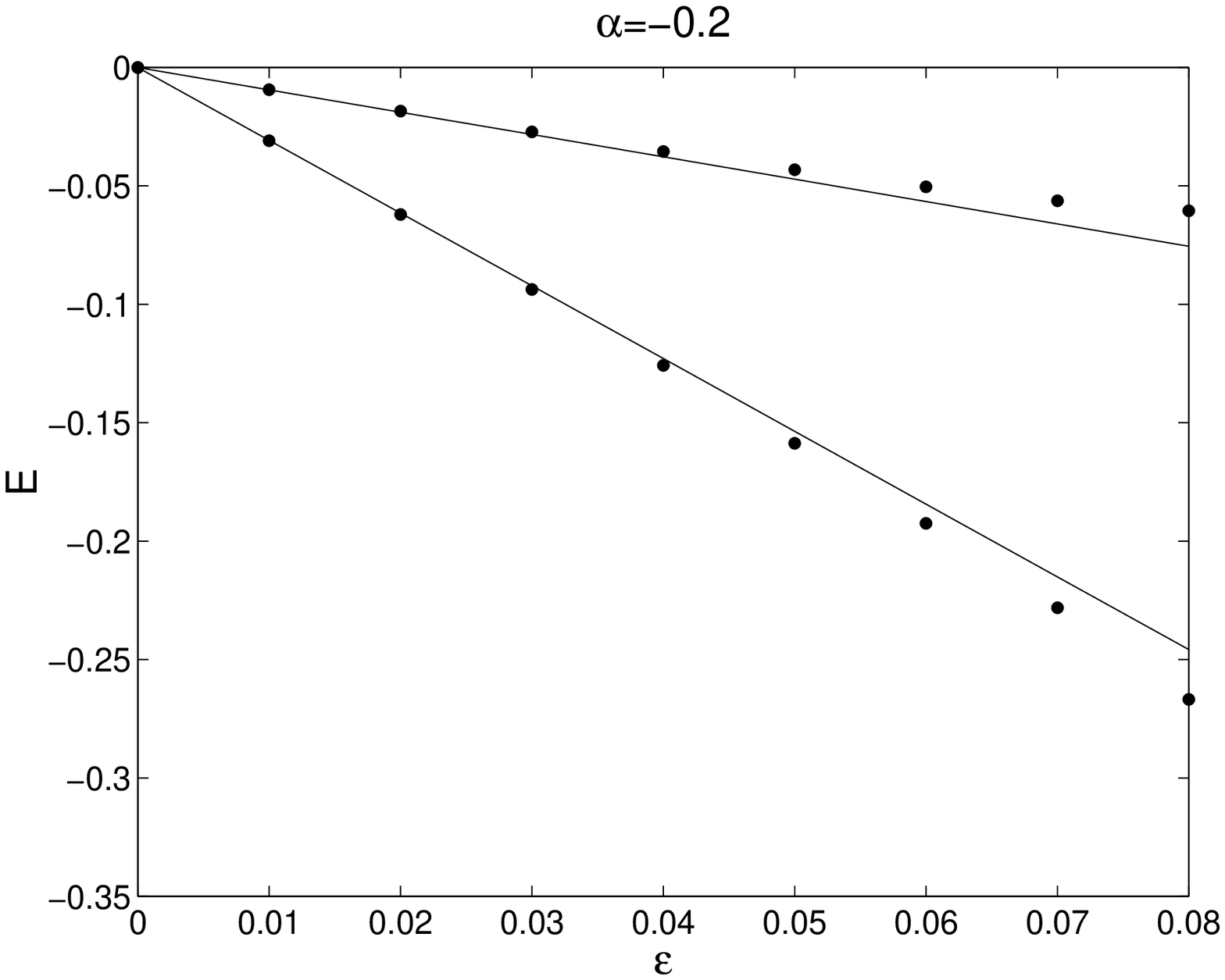} &
    \includegraphics[width=\middlefig]{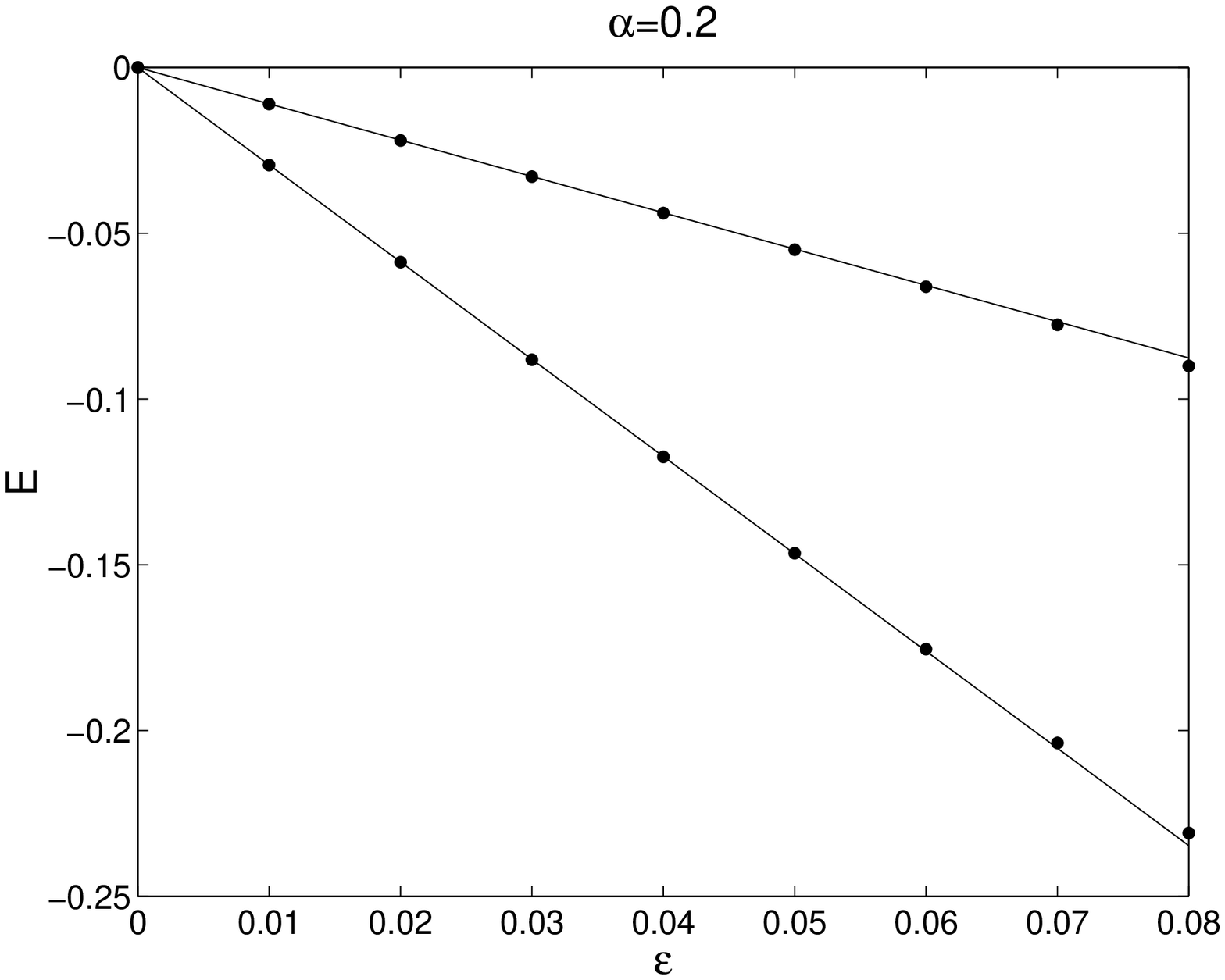}\\
\end{tabular}
\caption{Dependence of $E=\ee\lambda_1$ and $E=\ee\lambda_2$ with
respect to $\ee$ for two different values of $\alpha$.  (a,b)
correspond to in-phase 3-site breathers, (c,d) to out-of-phase
3-site breathers with the impurity at the center  and (e,f) to the
same as before but with the impurity at the edge. All the
solutions stand for a hard $\phi^4$ potential and
$\wb=1.3$. Points represent numerical solutions and lines analytical ones.}%
\label{fig:3sh}
\end{center}
\end{figure}

Figs. \ref{fig:3sm} and \ref{fig:3sh} show the dependence of
$E_1=\ee\la_1$ and $E_2=\ee\la_2$ with respect to $\ee$ calculated
numerically and analytically. The agreement between both
descriptions is again very good even for fairly large values of
the coupling constant.

\subsection{Inhomogeneity at the coupling constant}

\begin{figure}
\begin{center}
\begin{tabular}{cc}
    (a) & (b) \\
    \includegraphics[width=\middlefig]{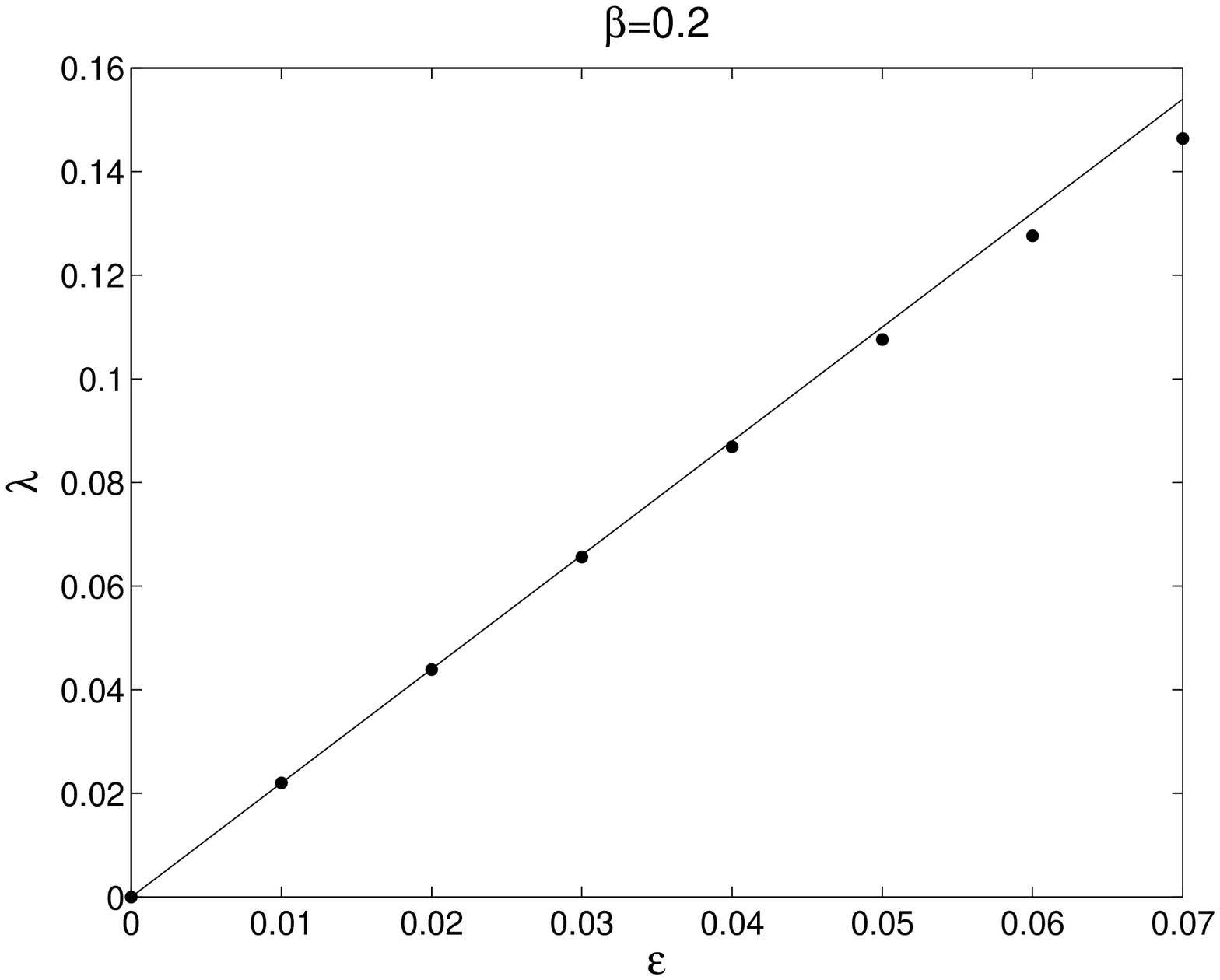} &
    \includegraphics[width=\middlefig]{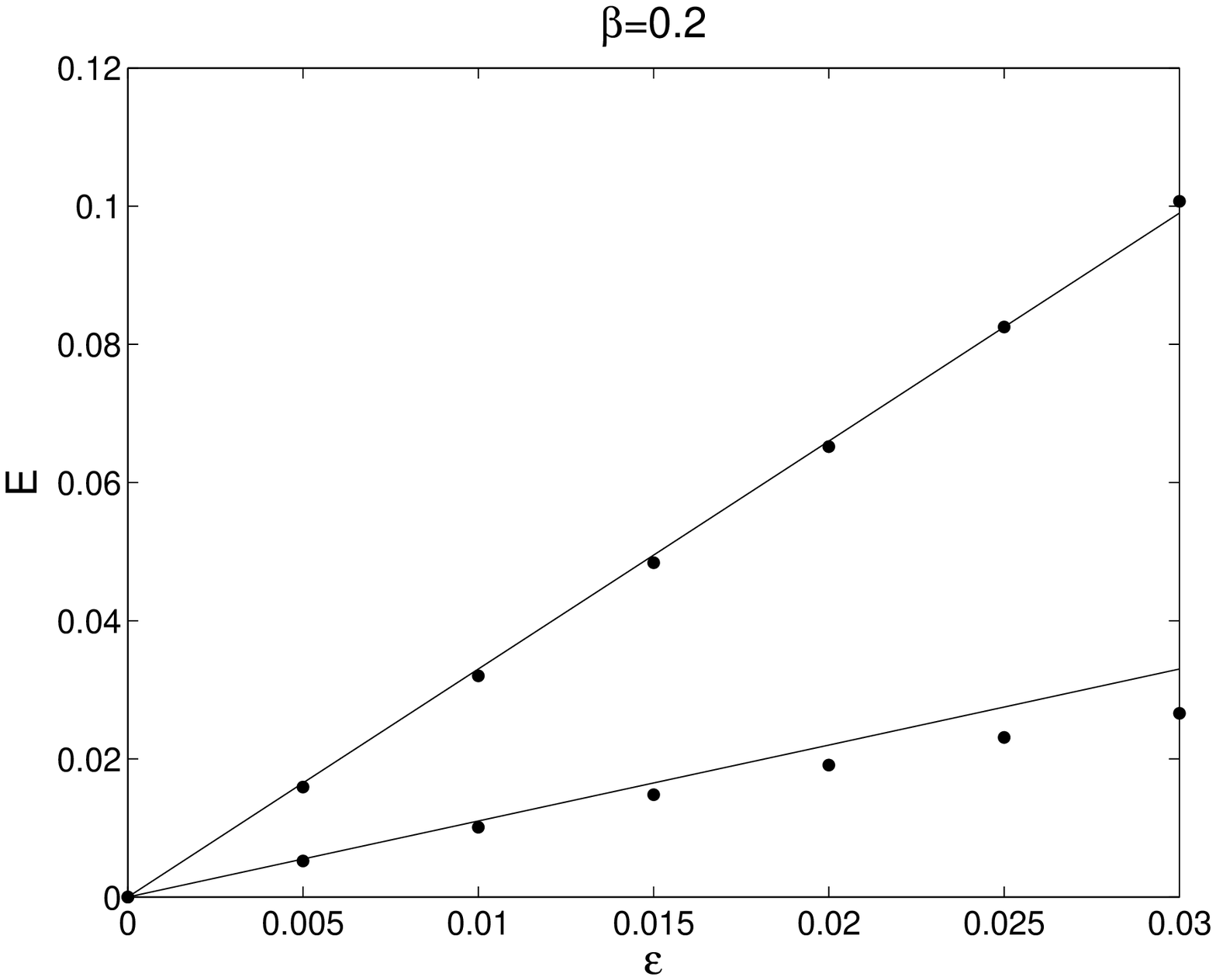}\\
\end{tabular}
\caption{Dependence of $E=\ee\lambda_1$ for a 2-site and a 3-site
breather with an impurity at the coupling located at the edge. All
the solutions stand for a Morse potential and $\wb=0.8$.
Points correspond to numerical solutions and lines to analytical ones.}%
\label{fig:coup}
\end{center}
\end{figure}

If the impurity is located at $n_0=2$, the relation
$Q=(\beta+2)/2\,Q_0$ is fulfilled again. The only way to obtain a
different relation is to suppose the impurity at an edge. The
following table summarizes the different cases:

\begin{center}

\begin{footnotesize}

\begin{tabular}{|c||c|c|}

\hline Code & $\la_1$ & $\la_2$ \\
\hline \hline%
1\ {\bf 1}\ 1 & $1+\beta/2$ & $3(1+\beta/2)$ \\%
\hline%
1\ 1\ {\bf 1} & $(\beta+4+\sqrt{\beta^2+4})/2$ & $(\beta+4-\sqrt{\beta^2+4})/2$ \\%
\hline%
-1\ {\bf 1}\ -1 & $-\g_0(1+\beta/2)$ & $-3\g_0(1+\beta/2)$ \\%
\hline%
-1\ 1\ {\bf -1} & $-\g_0(\beta+4+\sqrt{\beta^2+4})/2$ & $-\g_0(\beta+4+\sqrt{\beta^2+4})/2$ \\%
\hline%
{\bf 1}\ 1\ -1 &
$1-\g_0+(\beta+\sqrt{\g_0^2\beta^2+2\g_0(2\g_0+1)\beta+4(\g_0^2+\g_0+1)})/2$
&
$1-\g_0+(\beta-\sqrt{\g_0^2\beta^2+2\g_0(2\g_0+1)\beta+4(\g_0^2+\g_0+1)})/2$ \\%
\hline%
1\ {\bf 1}\ -1 & $(1+\beta/2)(1-\g_0+\sqrt{\g_0^2+\g_0+1})$ &
$(1+\beta/2)(1-\g_0+\sqrt{\g_0^2+\g_0+1})$
\\
\hline%
1\ 1\ {\bf -1} &
$1-\g_0+(\beta+\sqrt{\g_0^2\beta^2+2\g_0(2\g_0+1)\beta+4(\g_0^2+\g_0+1)})/2$
&
$1-\g_0+(\beta-\sqrt{\g_0^2\beta^2+2\g_0(2\g_0+1)\beta+4(\g_0^2+\g_0+1)})/2$ \\%
\\
\hline
\end{tabular}
\end{footnotesize}
\end{center}
 The signs of the eigenvalues are the same as in the on-site
inhomogeneity case. Fig.~\ref{fig:coup} compares these results
with the numerically obtained ones.

\section{Multibreathers and phonobreathers with an impurity} \label{sec:phono}

A multibreather is obtained from the anticontinuous limit when a
number of contiguous oscillators are excited. If all the
oscillators are excited it is called a phonobreather. Provided
that a multibreather does not include the oscillators at the
borders, the modified coupling matrix of a N-site multibreather is
identical to the one of a phonobreather with fixed ends.
Therefore, we can study both cases simultaneously. We cannot
obtain analytical expressions for the eigenvalues of the
perturbation matrixes, but we have been able to demonstrate their
stability properties.

 When the number of oscillator of a  multibreather increases, the
degree of the characteristic polynomial also increases and an
analytical evaluation of its roots is not possible. However, in
the case of a homogenous system, the eigenvalues can be calculated
because the eigenvectors equation of the system is equivalent to
the dynamical equations of a linear lattice of oscillators, as
shown in \cite{ACSA03}. Although this analogy might be suggested
in our case, the only information available for the eigenvalues
would be the dependence on the parameters of a localized
eigenvalues and only in an infinite lattice.

Thus, we can only obtain qualitative results about the spectrum of
the matrix $Q$ for a phonobreather (or N-site multibreather). In
particular, we  show that the stability properties of the system
with an impurity are the same as an homogeneous system. In order
to do that, we will make use of some congruence properties of
symmetric matrices \cite{Horn}.

Two symmetric matrices $A$ and $B$ are congruent if they can be
transformed each other through elementary transformations:
$B=PAP^t$. The inertia of a matrix is defined as
$\In(A)=\{i_{+}(A),i_{-}(A),i_{0}(A)\}$ where $i_{+}$, $i_{-}$,
$i_{0}$ denote, respectively, the number of positive, negative and
zero eigenvalues of $A$. Sylvester's inertia law establishes that
the inertia of two congruent matrices are the same
\cite{Sylvester}. In consequence, it is enough to diagonalize a
matrix using elementary transformations to obtain its inertia. The
diagonal matrix has the structure called first canonical form:

\begin{equation}
    D=\{\underbrace{1,\ldots,1}_{i_{+}\ \textrm{times}},
    \underbrace{-1,\ldots,-1}_{i_{-}\ \textrm{times}},
    \underbrace{0,\ldots,0}_{i_{0}\ \textrm{times}}
    \}
\end{equation}

We analyze below the stability properties of phonobreathers with
different boundary conditions and an inhomogeneity at the on-site
potential. Let us recall that, in the homogeneous case, the
inertia of $Q$ is, for an in-phase breather, $\In(Q)=\{N-1,0,1\}$
and, for an out-of-phase (staggered) breather,
$\In(Q)=\{0,N-1,1\}$, with $N$ being the number of system
particles. We will show that the inertia does not change when an
impurity is introduced.

Furthermore, the only multibreathers that can be stable are those
that vibrate in-phase or staggered, as demonstrated in \ref{ap:b}.

\subsection{Multibreathers and phonobreathers with free/fixed ends boundary conditions}
\label{subsec:phonoF}

If the impurity is supposed to be located at $n=c$, the elements
of the perturbation matrix $Q^{F0}$ for an in-phase multibreather,
can be written as:

\begin{eqnarray}
\fl    Q^{F0}_{nm} = 2\d_{n,m}[1+(\h\p-1)\d_{n,c}+ (\h/\p-1)/2
    \d_{|n-c|,1}-(\d_{n,1}+\d_{n,N})/2]- \nonumber \\
   \fl - \d_{n,m+1}[1-(\h+1)(\d_{n,c}+\d_{n,c+1})]
    -\d_{n,m-1}[1-(\h+1)(\d_{n,c}+\d_{n,c-1})].
\end{eqnarray}
This matrix is transformed into a diagonal matrix through
$D=PQP^t$, where

\begin{eqnarray}
\fl    P_{nm} = [\d_{[\frac{m}{n}],0}+\d_{n,m}]
    [1+\d_{[\frac{m}{c}],0}\d_{n,c-1}(\sqrt{\p/\h}-1)+\d_{m,c}\d_{[\frac{n}{c}],0}/\p+
    \nonumber \\
    \lo{+} \d_{n,c-1}((\sqrt{\p/\h}-1)\d_{[\frac{m}{c}],0}+(1/\sqrt{\p\h}-1)\d_{m,c})],
\end{eqnarray}
where $[\cdot]$ denotes the integer part.
$D_{nm}=\d_{n,m}(1-\d_{n,N})$. In consequence,
$\In(Q^{F0})=\{N-1,0,1\}$ and all of the eigenvalues are positive
except for one which is zero.

The perturbation matrix for the out-of-phase breather is given by:
\begin{eqnarray}
 \fl   Q^{F\pi}_{nm} = -2\g_0\d_{n,m}[1+(\g\p-1)\d_{n,c}+ (\g/\p-1)/2
    \d_{|n-c|,1}-(\d_{n,1}+\d_{n,N})/2]- \nonumber \\
\fl    + \g_0\d_{n,m+1}[1-(\g_0+1)(\d_{n,c}+\d_{n,c+1})]
    + \g_0\d_{n,m-1}[1-(\g/\g_0+1)(\d_{n,c}+\d_{n,c-1})],
\end{eqnarray}
which transform into the diagonal matrix
$D_{nm}=-\d_{n,m}(1-\d_{n,N})$ through the transformation matrix
\begin{eqnarray}
\fl     P_{nm} = [\d_{[\frac{m}{n}],0}+\d_{n,m}]
    [1+\d_{[\frac{m}{c}],0}\d_{n,c-1}(\sqrt{\p/\g}-1)+\d_{m,c}\d_{[\frac{n}{c}],0}/\p+
    \nonumber \\
   \lo +
    \d_{n,c-1}((\sqrt{\p/\g}-1)\d_{[\frac{m}{c}],0}+(1/\sqrt{\p\g}-1)\d_{m,c})].
    \end{eqnarray}

The inertia of the stability matrix is given by
$\In(Q^{F\pi})=\{0,N-1,1\}$ and, in consequence, it has all its
eigenvalues negatives except for the null one.

\begin{figure}
\begin{center}
\begin{tabular}{cc}
    (a) & (b) \\
    \includegraphics[width=\middlefig]{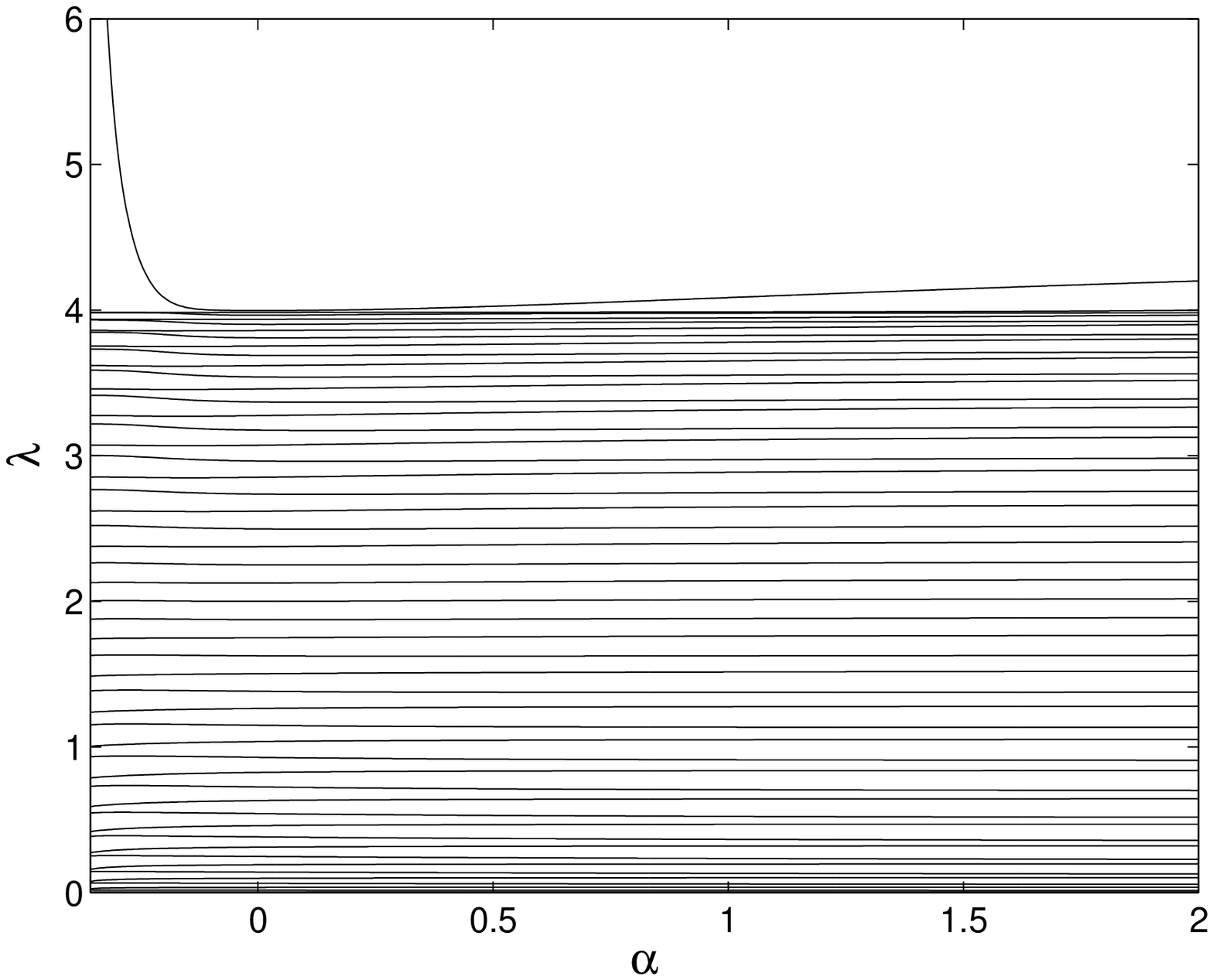} &
    \includegraphics[width=\middlefig]{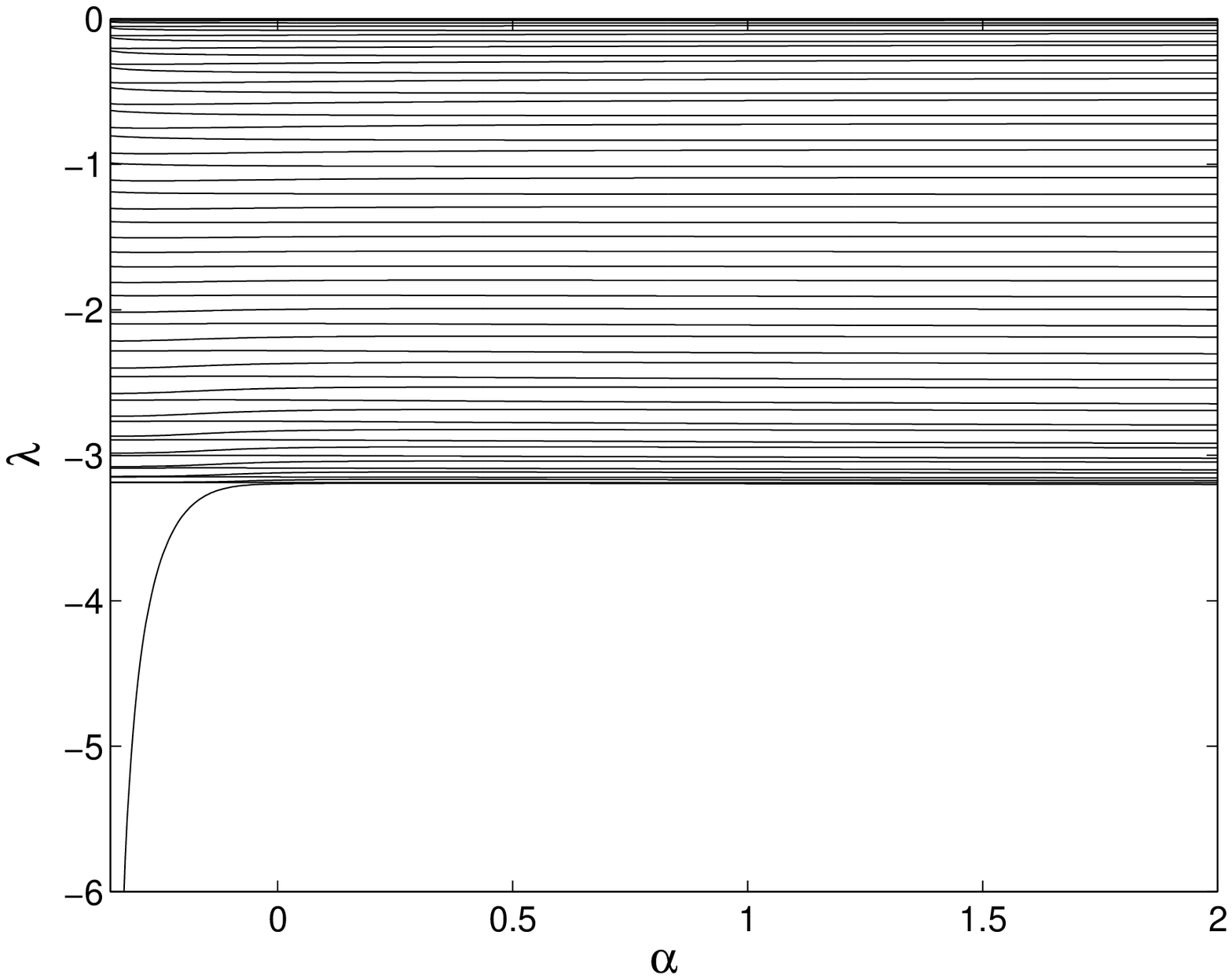}\\
    \multicolumn{2}{c}{(c)} \\
    \multicolumn{2}{c}{\includegraphics[width=\middlefig]{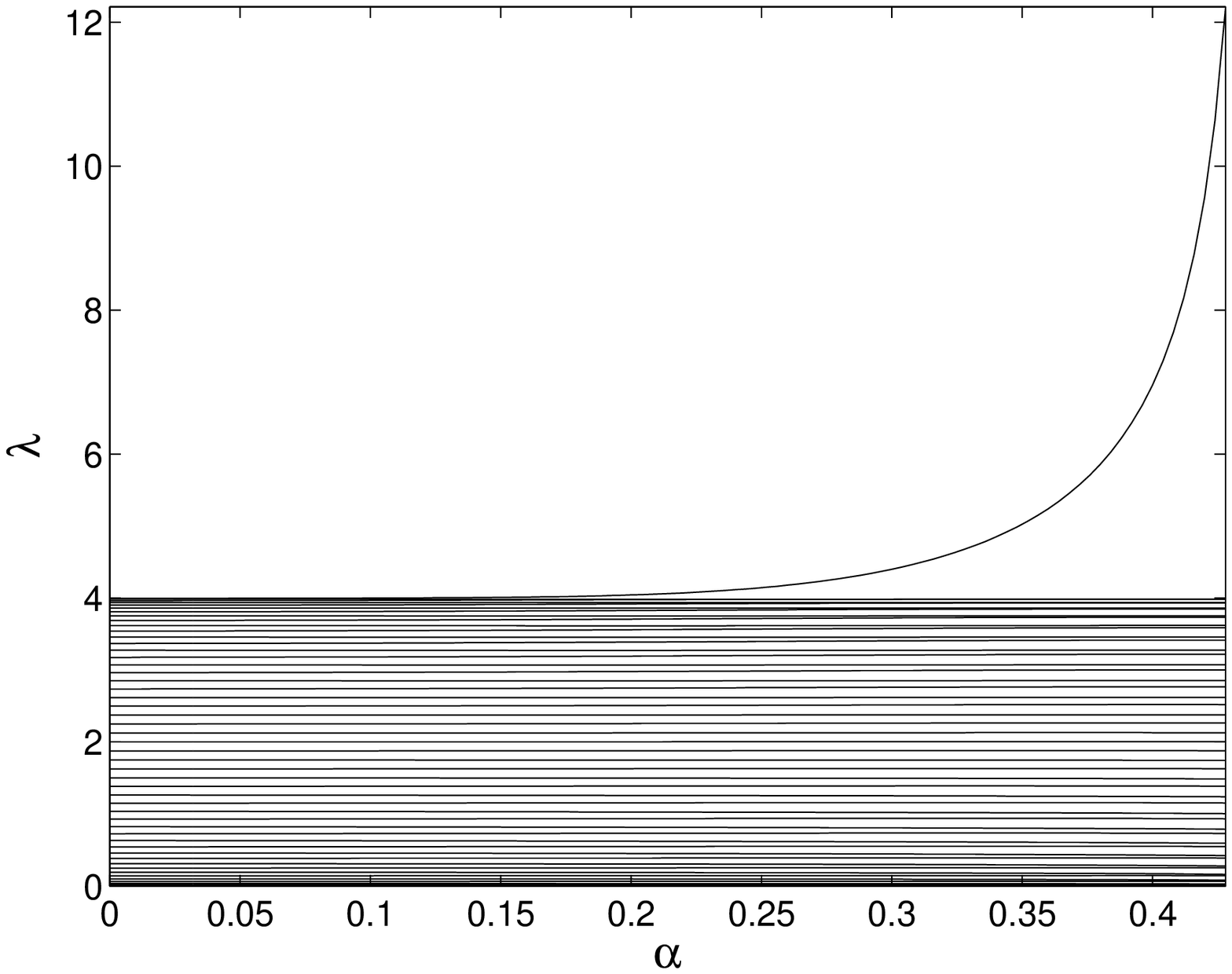}}\\
\end{tabular}
\caption{Dependence of the eigenvalues of the perturbation matrix
with respect to the impurity parameter $\alpha$ for a
phonobreather with free/fixed ends boundary condition. (a) and (b)
corresponds to the Morse potential and $\wb=0.8$ while (c) to a
hard $\phi^4$ potential. (a) and (c) corresponds to a in-phase
breather and (b) to a staggered one.}%
\label{fig:phonof}
\end{center}
\end{figure}

Figure \ref{fig:phonof} shows the numerical values of the
eigenvalues and confirms the results stated previously.

\subsection{Periodic boundary conditions} \label{sec:phonoP}

The diagonalization of $Q$ is now very difficult. We will make use
instead of a theorem due to Weyl \cite{Horn} to obtain the inertia
of the matrix. This theorem establishes that, if the eigenvalues
of a matrix are arranged in the following order:
$\la_1\leq\la_2\leq\ldots\leq\la_N$ then, the k-th eigenvalue of
the sum of two matrices holds:

\begin{equation} \label{eq:Weyl}
    \la_k(A)+\la_1(B)\leq\la_k(A+B)\leq\la_k(A)+\la_N(B).
\end{equation}

This theorem can be particularized for our case if we denote as
$Q^{P0}$ and $Q^{P\pi}$ to the periodic boundary conditions
matrices and define $Q^P=Q^F+\Delta$, where $Q^F$ is the
perturbation matrix with free boundary conditions. By making
$A\equiv Q^F$ and $B\equiv\Delta$    we can apply Weyl's theorem.

For a phonobreather in-phase,
\begin{eqnarray}
 \fl   Q^{P0}_{nm} = 2\d_{n,m}[1+(\h\p-1)\d_{n,c}+(\h/\p-1)/2
    \d_{|n-c|,1}]-(\d_{n,1}\d_{m,N}+\d_{n,N}\d_{m,1})- \nonumber \\
    \fl - \d_{n,m+1}[1-(\h+1)(\d_{n,c}+\d_{n,c+1})]
    -\d_{n,m-1}[1-(\h+1)(\d_{n,c}+\d_{n,c-1})],
\end{eqnarray}
and, in consequence,
\begin{equation}
    \Delta_{n,m}=\d_{n,m}(\d_{n,1}+\d_{n,N})-(\d_{n,1}\d_{m,N}+\d_{n,N}\d_{m,1}).
\end{equation}

The spectrum of $\Delta$ is given by
$\mathrm{spec}(\Delta)=\{0,2\}$ where $\la_1(\Delta)=0$ has
multiplicity $N-1$ and $\la_N(\Delta)=2$ has multiplicity 1. Thus,
particularizing (\ref{eq:Weyl}) for $k=1$:

\begin{equation}
    \la_1(Q^{F0})+\la_1(\Delta)\leq\la_1(Q^{P0})\leq\la_1(Q^{F0})+\la_N(\Delta).
\end{equation}

As $\la_1(Q^{F0})=0$, $\la_1(Q^{F0})\in[0,2]$. If $k=2$ is taken
in (\ref{eq:Weyl}),
$\la_2(Q^{P0})\geq\la_2(Q^{F0})+\la_1(\Delta)$. As
$\la_2(Q^{F0})>0$ then $\la_2(Q^{P0})>0$. In consequence, $Q^{P0}$
has $N-1$ positive eigenvalues and one which is greater or equal
to zero.

The steps for an out-of-phase multibreather are similar to the
in-phase case, except that $k=N$ is taken in (\ref{eq:Weyl}).
Thus, it is straightforward to show that $Q^{P\pi}$ has all its
eigenvalues negatives except for one of them which is smaller or
equal to zero.

\begin{figure}
\begin{center}
\begin{tabular}{cc}
    (a) & (b) \\
    \includegraphics[width=\middlefig]{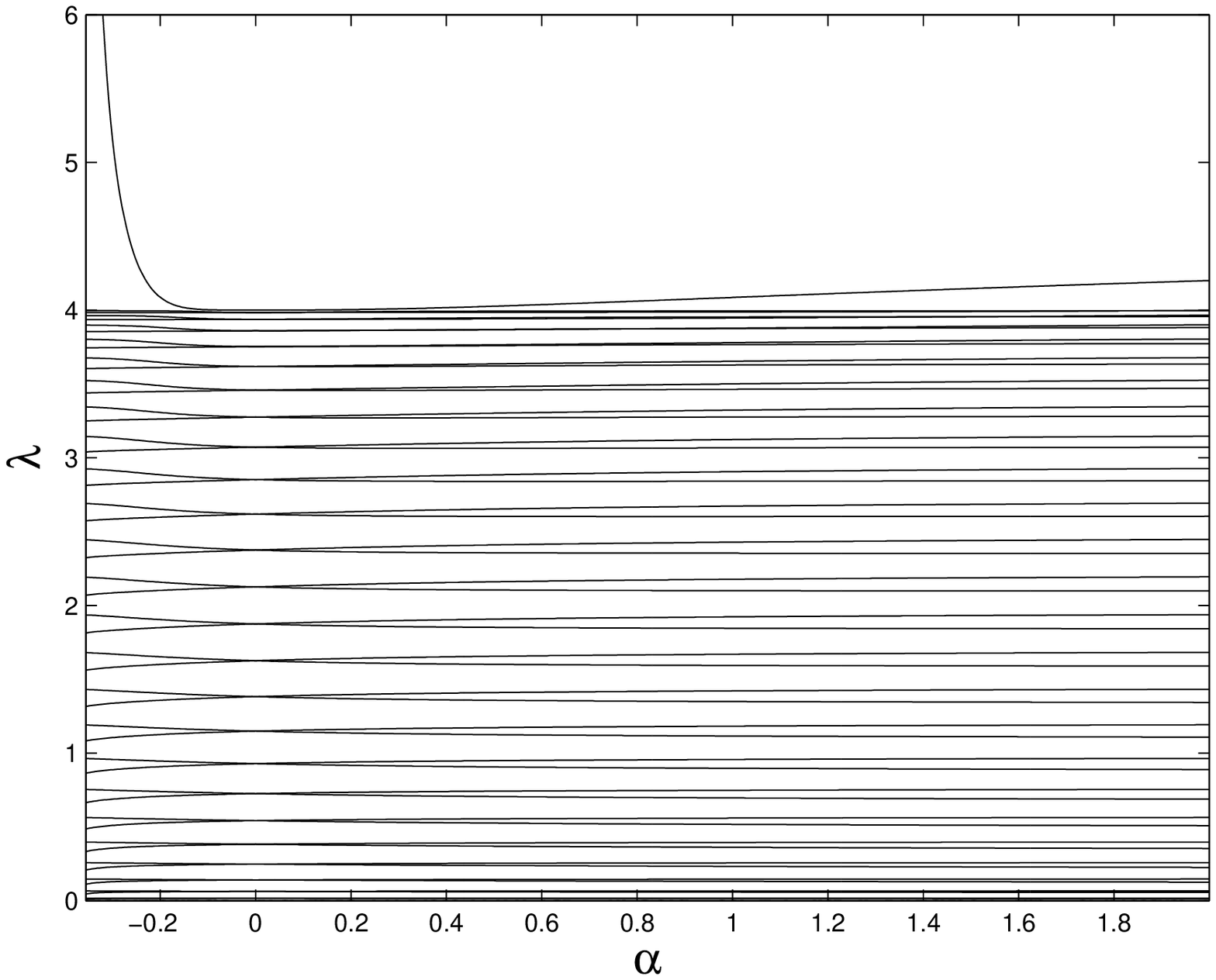} &
    \includegraphics[width=\middlefig]{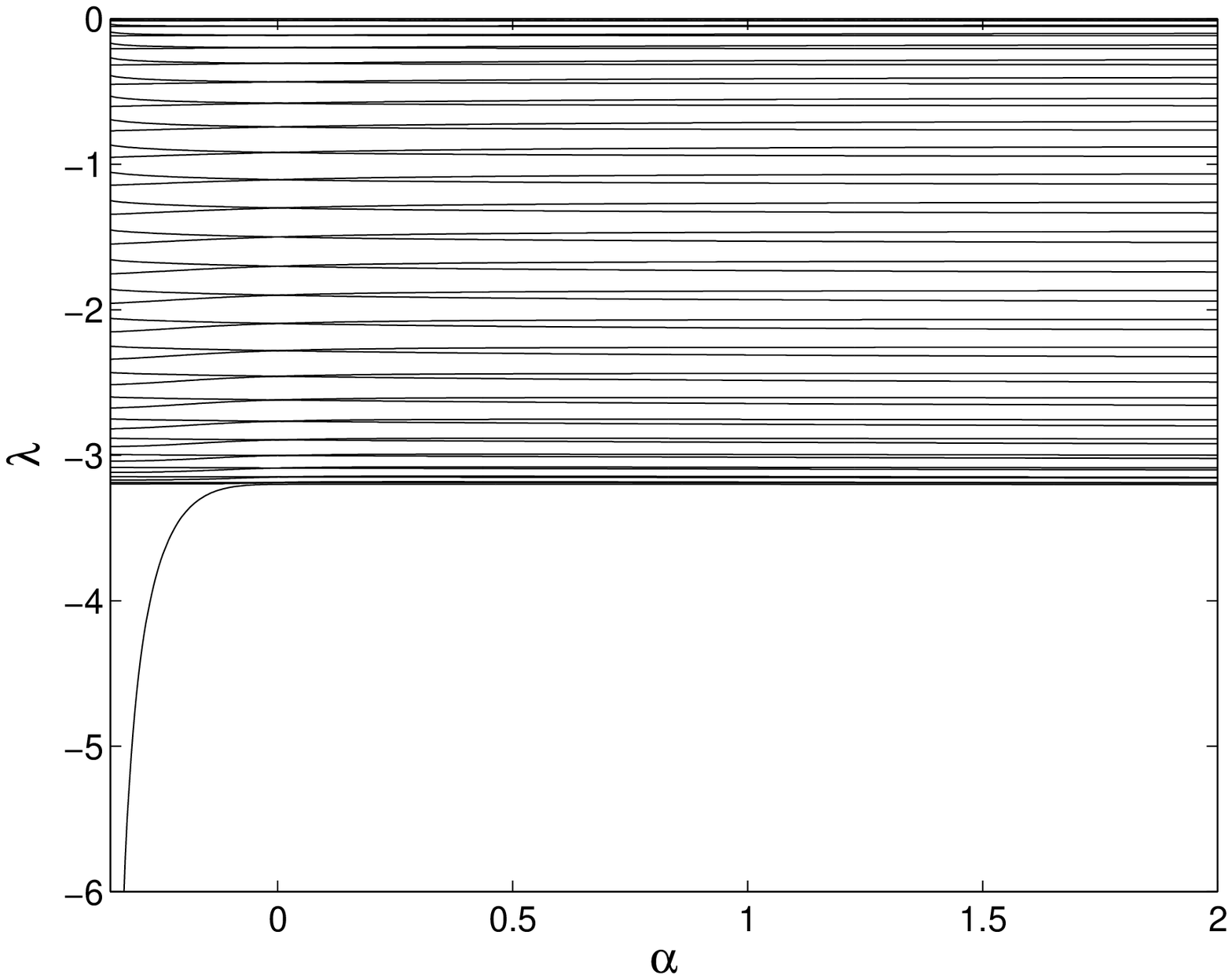}\\
    \multicolumn{2}{c}{(c)} \\
    \multicolumn{2}{c}{\includegraphics[width=\middlefig]{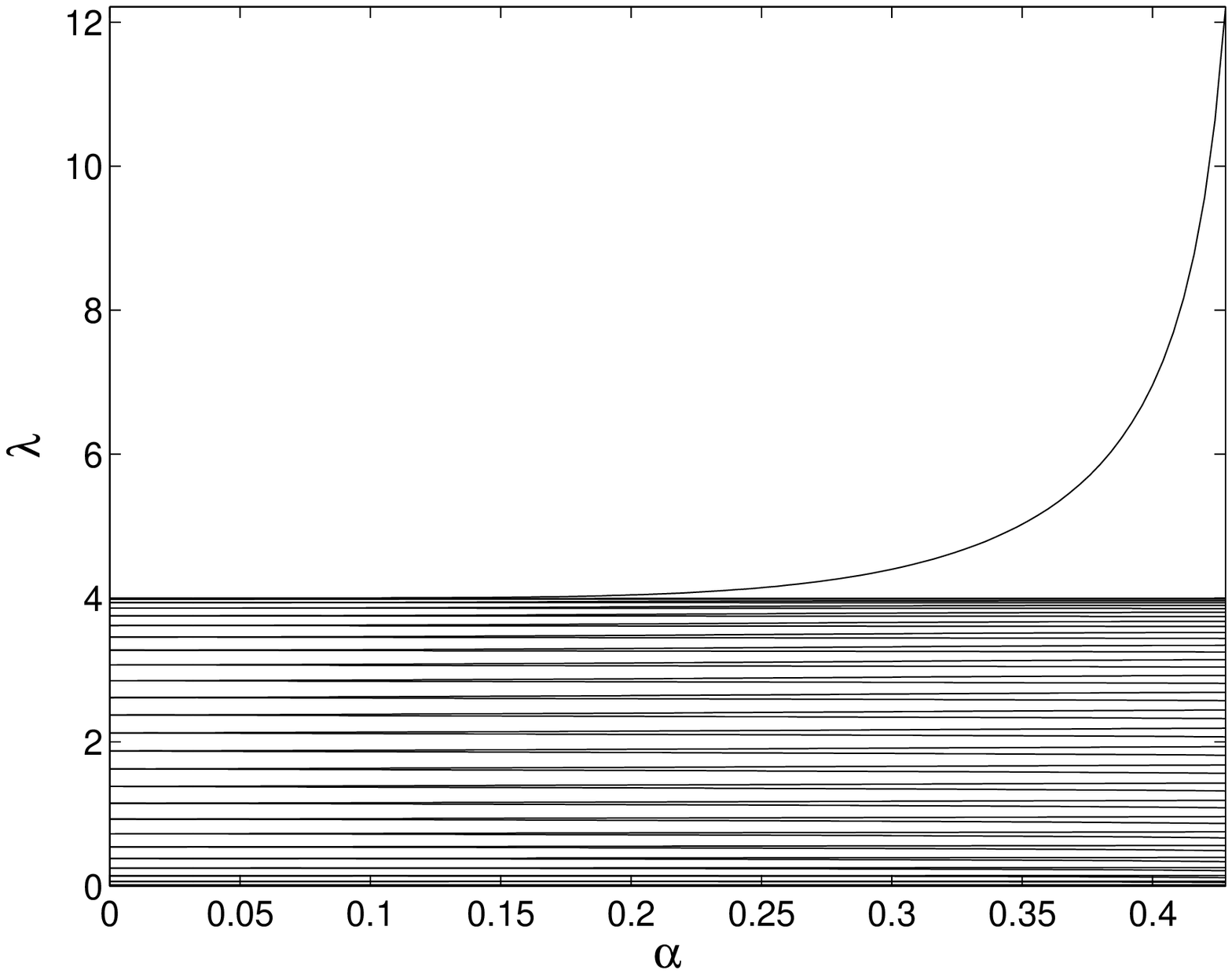}}\\
\end{tabular}
\caption{Dependence of the eigenvalues of the perturbation matrix
with respect to the impurity parameter $\alpha$ for a
phonobreather with periodic boundary condition. (a) and (b)
corresponds to Morse potential and $\wb=0.8$ while (c) to a hard
$\phi^4$ potential. (a) and (c) corresponds to a in-phase
breather and (b) to a staggered one.}%
\label{fig:phonop}
\end{center}
\end{figure}

Figure \ref{fig:phonop} shows the numerical values of the
eigenvalues and confirms the previously stated results.

\begin{figure}
\begin{center}
\begin{tabular}{cc}
    (a) & (b) \\
    \includegraphics[width=\middlefig]{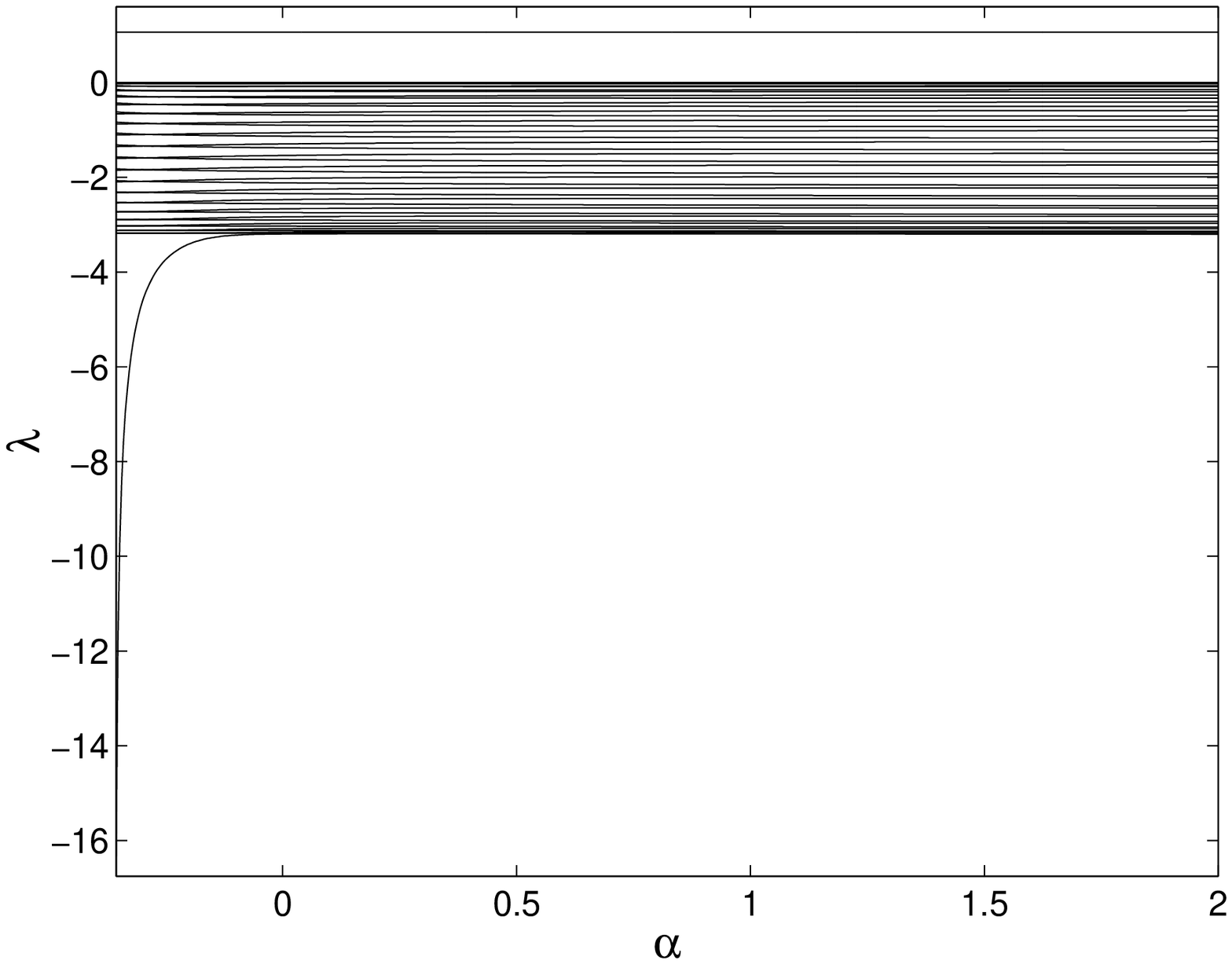} &
    \includegraphics[width=\middlefig]{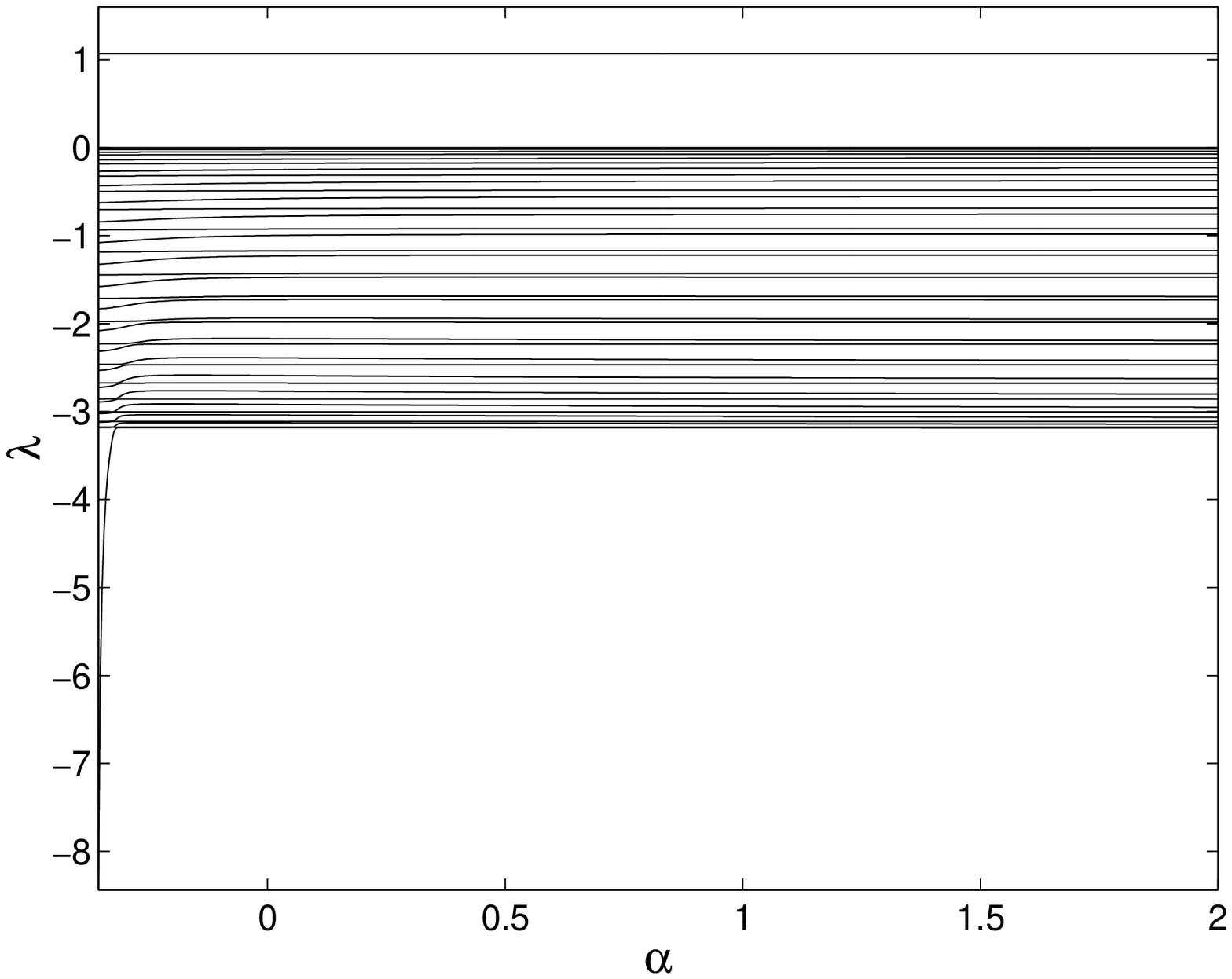}\\
\end{tabular}
\caption{Dependence of the eigenvalues of the perturbation matrix with
respect to the impurity parameter $\alpha$ for a Morse potential
and $\wb=0.8$ in the case of staggered (a) phonobreathers and (b) dark breathers
with periodic boundary conditions. The existence of an isolated positive eigenvalue
$\lambda=4\g_0/3\approx1.07$ is observed.}%
\label{fig:parity}
\end{center}
\end{figure}

Let us remark that for $N$ odd there appear, in a similar fashion
to the homogeneous case, parity instabilities due to the breaking
of the breather pattern. Figure \ref{fig:parity} shows the
dependence of the eigenvalues of $Q$ with respect to $\alpha$ in
this case. It can be observed the existence of a positive
eigenvalue of constant value. It was demonstrated in \cite{Tesis}
that in a homogeneous lattice, this eigenvalue is $4\g_0/3$ in an
infinite lattice. This value does not change when an impurity is
introduced as it is due to a border effect.

\subsection{Inhomogeneity at the coupling constant}

In this case, the perturbation matrix is equivalent to the one
corresponding to an inhomogeneity implemented at the on-site
potential with the changes $\p=1$ and $\eta=\gamma=1+\beta/2$. In
consequence, the results are qualitatively the same in both cases.

\section{Dark breathers with an impurity}

A particular case of multibreathers in which all the particles of
the lattice are excited except one is called a dark breather
\cite{AACR02,MJKA02}. In the case that there are $p$ adjacent
sites with no vibration, we deal with a p-site dark breather. For
the sake of simplicity we will consider 1-site dark breathers as
the generalization to p-site dark breathers is straightforward.

It is worth noticing that if the impurity is located at the dark
site, the perturbation matrix turns into the homogeneous one. In
consequence, the only way to introduce a modification in the
perturbation matrix is to suppose the impurity adjacent to the
dark site.

For an in-phase breather, the perturbation matrix is given by
(notice that the $c+1$ row and column are removed):
\begin{eqnarray}
 \fl   Q^{D0}_{nm} = 2\d_{n,m}[1+(\h\p-1)\d_{n,c}+ (\h/\p-1)/2
    \d_{n,c-1}+(b-1)(\d_{n,1}+\d_{n,N})/2]- \nonumber \\
\fl    - \d_{n,m+1}[1-(\h+1)\d_{n,c-1}]
    -\d_{n,m-1}[1-(\h+1)\d_{n,c-1}]-b(\d_{1,N}+\delta_{1,N}),
\end{eqnarray}
where $b=0$ for fixed/free ends and $b=1$ for periodic boundary
conditions. If $b=0$ the perturbation matrix is block diagonal,
being the lower block the matrix of a phonobreather with
fixed/free ends boundary conditions and $N-(c+1)$ particles. The
upper block is a matrix $Q_1$ with positive eigenvalues except for
a null one. In consequence, the inertia of the perturbation matrix
for $b=0$ is $\{N-2,0,2\}$. The existence of two zero eigenvalues
makes it impossible to determine the stability through the theorem
as, to first order of the perturbation, the degeneration is not
raised. This result was also obtained in the homogeneous case.

For $b=1$, i.e. periodic boundary conditions, the perturbation
matrix can be transformed into $Q_1$ through row/columns
interchange. This transformation lets the characteristic
polynomial invariant. In consequence, the matrix has $N-1$
positive eigenvalues and a null one.

The perturbation matrix of a staggered dark breather is
\begin{eqnarray}
 \fl   Q^{D\pi}_{nm} = -2\g_0\d_{n,m}[1+(\g\p-1)\d_{n,c}+ (\g/\p-1)/2
    \d_{n,c-1}+(b-1)(\d_{n,1}+\d_{n,N})/2]- \nonumber \\
 \lo - \g_0\d_{n,m+1}[1-(\g/\g_0+1)\d_{n,c-1}]\nonumber \\
\lo
-\g_0\d_{n,m-1}[1-(\g/g_0+1)\d_{n,c-1}]-b\g_0(\d_{1,N}+\delta_{1,N}).
\end{eqnarray}

It can be shown straightforwardly that
$\In(Q^{D\pi}(b=0))=\{0,N-2,2\}$ and
$\In(Q^{D\pi}(b=1))=\{0,N-1,1\}$. In the case of $N$ odd and
$b=1$, parity instabilities also occur (see Figure
\ref{fig:parity}).

\begin{figure}
\begin{center}
\begin{tabular}{cc}
    (a) & (b) \\
    \includegraphics[width=\middlefig]{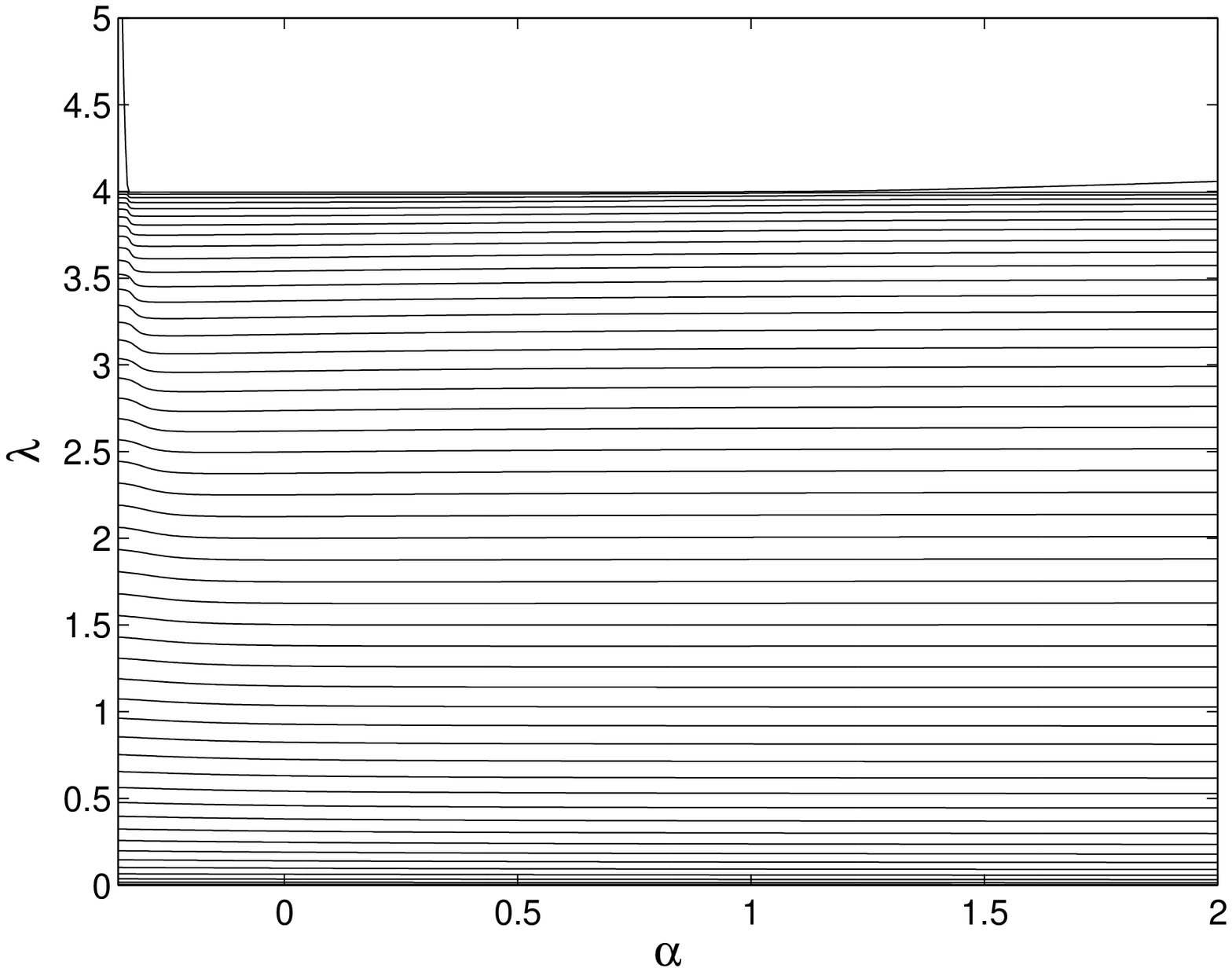} &
    \includegraphics[width=\middlefig]{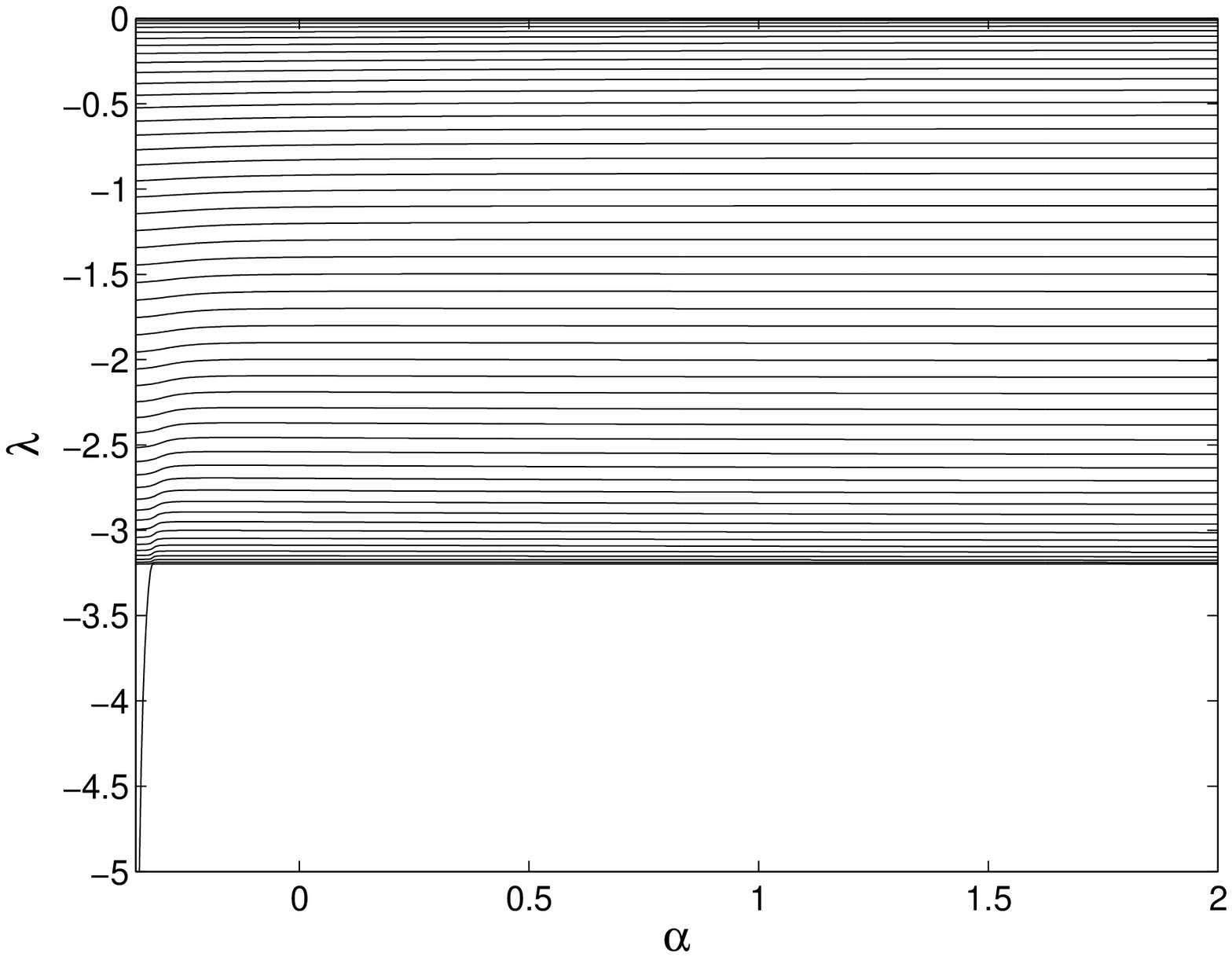}\\
    \multicolumn{2}{c}{(c)} \\
    \multicolumn{2}{c}{\includegraphics[width=\middlefig]{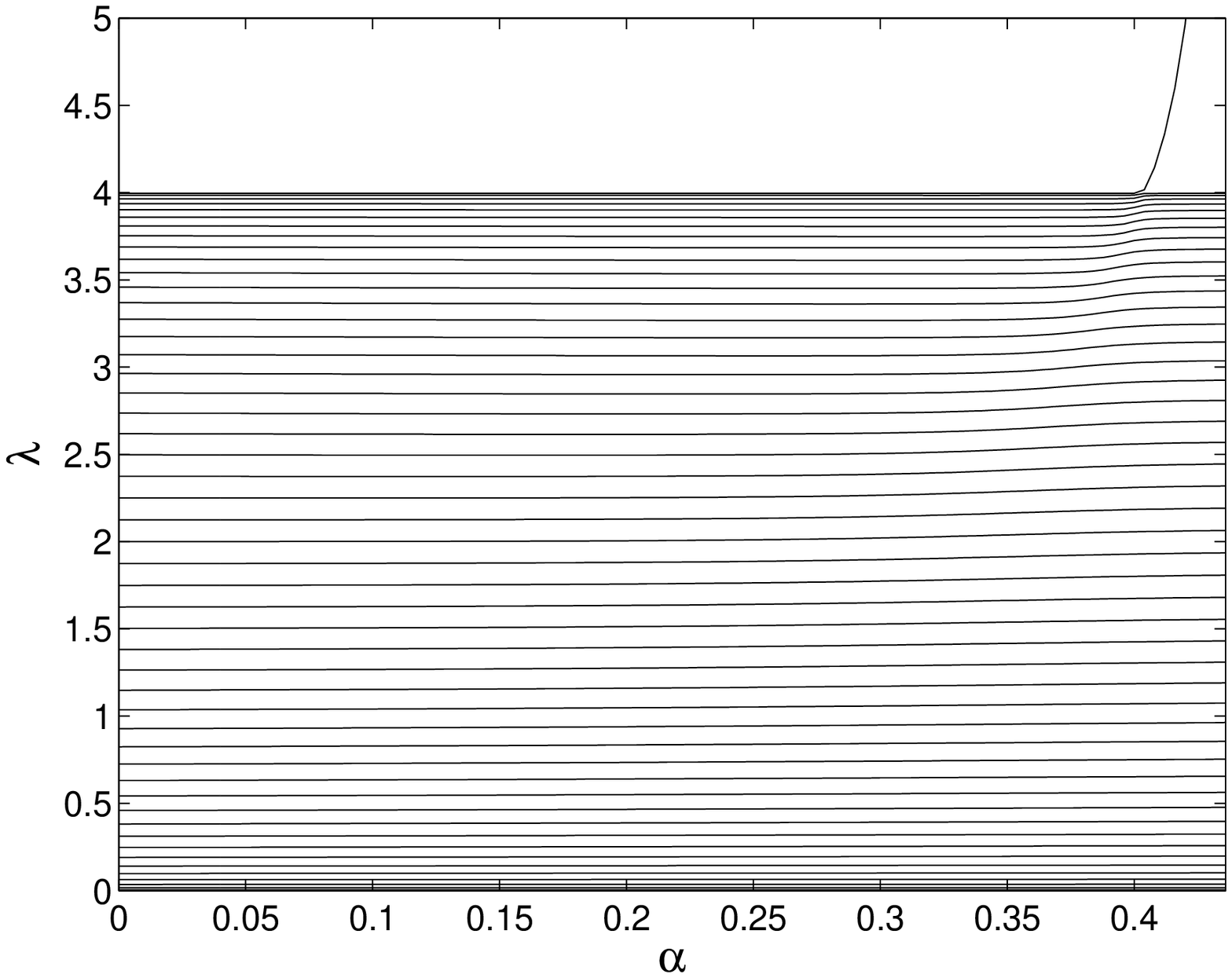}}\\
\end{tabular}
\caption{Dependence of the eigenvalues of the perturbation matrix
with respect to the impurity parameter $\alpha$ for a dark
breather with free/fixed ends boundary condition. (a) and (b)
corresponds to Morse potential and $\wb=0.8$ while (c) to a hard
$\phi^4$ potential. (a) and (c) corresponds to a in-phase
breather and (b) to a staggered one.}%
\label{fig:dark}
\end{center}
\end{figure}

Figure \ref{fig:dark} shows the eigenvalues for a dark breather
with periodic boundary conditions.

\section{Conclusions}

The {\em Multibreathers Stability Theorem}~\cite{ACSA03} provides
with a method for obtaining the perturbation matrix $Q$ from the
anticontinuous limit of any multibreather at low coupling in
Klein-Gordon systems. The stability depends on the signs of the
eigenvalues of $Q$, the hardness/softnes of the on-site potential
and the sign of the coupling parameter (i.e., the attractiveness
or repulsiveness of the coupling). In this paper we apply the MST
to systems with an impurity either at the on-site potential or the
coupling constants, the first case being equivalent to an impurity
at the masses.

Using analytical methods we have been able to obtain the
eigenvalues of $Q$ for 2-site and 3-site breathers and compare
them with numerical results showing an excellent agreement. For
larger multibreathers, phonobreathers and dark breathers we have
obtained analytically the signs of the $Q$-eigenvalues and checked
them numerically. In all cases, although the values of the
eigenvalues change, the stability properties of the system with an
impurity coincides with the ones of the homogeneous system.

The necessity of either an in-phase pattern or an out-of-phase one
for the stability of a multibreather in the homogenous system, as
it was conjectured in Ref.~\cite{ACSA03}, has also been
demonstrated. 

\appendix
\section{Analytical calculation of parameters} \label{ap:a}

The aim of this appendix is to calculate analytically the
parameters $\h_{nm}$, $\g_{nm}$, $\p_{nm}$ defined in Section
\ref{sec:stab} as:
\begin{equation}\label{eq:hgpa}
 \fl   \h_{nm}=\frac{J_{nm}}{\sqrt{J_{nn}J_{mm}}}>0, \qquad
    \g_{nm}=-\frac{J'_{nm}}{\sqrt{J_{nn}J_{mm}}}>0, \qquad
    \p_{nm}=\sqrt{\frac{J_{mm}}{J_{nn}}}>0,
\end{equation}
with
\begin{equation}
 \fl   J_{nm}=\int_{-T/2}^{T/2}\,\dot{u}_n(t)\,\dot{u}_m(t)\,\dd t\,,
    \qquad
    J'_{nm}=\int_{-T/2}^{T/2}\,\dot{u}_n(t)\,\dot{u}_m(t+T/2)\,\dd
    t\,.
\end{equation}

Analytical calculation of these integrals is rather difficult. In
order to overcome this difficulty, solutions $u_n(t)$ are
expressed as a Fourier series expansion:
\begin{equation}
    u_n(t)=z_{0,n}+2\sum_{k=1}^{\infty}z_{k,n}\cos(k\wt),
\end{equation}
and the integrals turn into series:
\begin{equation}
\fl    J_{nm}=4\pi\wb\sum_{k=1}^{\infty}k^2z_{k,n}z_{k,m},
    \qquad
    J'_{nm}=4\pi\wb\sum_{k=1}^{\infty}(-1)^kk^2z_{k,n}z_{k,m}.
\end{equation}

In consequence, the theorem parameters can be expressed as a
function of the Fourier coefficients of the solution. In the case
of the Morse potential, the sum can be performed and a closed
expression is obtained.

\subsection{Morse potential}

The Morse potential has the expression
$V(u_n)=(1+\alpha_n)(\exp(-u_n)-1)^2/2$, and the orbits of the
equation $\ddot u_n+V'(u_n)=0$ are given by \cite{ACSA03}:
\begin{equation}\label{xt}
    u_n(t)=\log\frac{1\mp\sqrt{1-(\wb/\wn)^2}\cos\wt}{(\wb/\wn)^2}\,,
\end{equation}
The minus sign corresponds to $u_n(0)>0$ and the plus sign to
$u_n(0)<0$ with $\wn=\sqrt{1+\alpha_n}$. The Fourier coefficients
are:
\begin{equation}
    z_{0,n}=\log\frac{\wn+\wb}{2\wb^2} \quad;\quad
    z_{k,n}=-\,\frac{\rho_k}{k}\left(\frac{\wn-\wb}
    {\wn+\wb}\right)^{k/2}\,,
\end{equation}
with $\rho_k=(-1)^{k+1}$ if $u_n(0)>0$ and $\rho_k=-1$ if
$u_n(0)<0$. Thus,
\begin{equation}
    J_{nm}=4\pi\wb\sum_{k=1}^{\infty}
    \left[\frac{(\wn-\wb)(\wm-\wb)}{(\wn+\wb)(\wm+\wb)}\right]^{k/2},
\end{equation}
\begin{equation}
    J'_{nm}=4\pi\wb\sum_{k=1}^{\infty}
    \left[-\frac{(\wn-\wb)(\wm-\wb)}{(\wn+\wb)(\wm+\wb)}\right]^{k/2}.
\end{equation}
The sums are geometric series whose ratios have absolute values
smaller than one. In consequence, they can be easily summed and
the resulting values of $J$ and $J'$ are:

\begin{equation}
    J_{nm}=4\pi\wb\frac{\sqrt{(\wn-\wb)(\wm-\wb)}}
    {\sqrt{(\wn+\wb)(\wm+\wb)}-\sqrt{(\wn-\wb)(\wm-\wb)}},
\end{equation}

\begin{equation}
    J'_{nm}=4\pi\wb\frac{\sqrt{(\wn-\wb)(\wm-\wb)}}
    {\sqrt{(\wn+\wb)(\wm+\wb)}+\sqrt{(\wn-\wb)(\wm-\wb)}}.
\end{equation}
Then, applying equation (\ref{eq:hgpa}), the theorem parameters
are:

\begin{equation}\label{eq:hmorse}
    \h_{nm}=\frac{2\wb}{\sqrt{(\wn+\wb)(\wm+\wb)}-\sqrt{(\wn-\wb)(\wm-\wb)}},
\end{equation}

\begin{equation}\label{eq:gmorse}
    \g_{nm}=\frac{2\wb}{\sqrt{(\wn+\wb)(\wm+\wb)}+\sqrt{(\wn-\wb)(\wm-\wb)}},
\end{equation}

\begin{equation}\label{eq:pmorse}
    \p_{nm}=\sqrt{\frac{\wm-\wb}{\wn-\wb}}.
\end{equation}

\subsection{$\phi^4$ potential}

This potential is given by $V(u_n)=(1+\alpha_n)u_n^2/2-su_n^4/4$,
with $s=\pm1$. The sign of $s$ allow us to define two different
cases:

\subsubsection{Hard $\phi^4$ potential. $s=+1$.}

The orbit of a hard oscillator is given by:
\begin{eqnarray}
\fl u_n(t)
=\pm\wn\sqrt{\frac{2\k_n^2}{1-2\k_n^2}}\cn\left(\frac{\wn
t}{\sqrt{1-2\k_n^2}},\k_n\right)\nonumber \\ \lo =
    \pm\wn\sqrt{\frac{2\k_n^2}{1-2\k_n^2}}\cn\left(\frac{2K(\k_n)}{\pi}\wt,\k_n\right),
\end{eqnarray}
where $\cn$ is a Jacobi elliptic function of modulus $\k_n$ and
$K(\k_n)$ is the complete elliptic integral of the first kind
defined as:
\begin{equation}
    K(\k)=\int_{0}^{\pi/2}\,\frac{\dd x}{\sqrt{1-\k^2\sin^2x}}\,.
\end{equation}
The breather frequency $\wb$ is related to the modulus $\k_n$
through:
\begin{equation}
    \wb=\frac{\pi\wn}{2\sqrt{1-2\k_n^2}K(\k_n)}.
\end{equation}
The elliptic function can be expanded into a Fourier series and it
is obtained \cite{Abram}:
\begin{equation}
    z_{2\nu+1,n}=\pm\frac{\pi}{K(\k_n)}\sqrt{\frac{2}{1-2\k_n^2}}\,
    \frac{q_n^{\nu+1/2}}{1+q_n^{2\nu+1}}, \qquad \nu=0,1,2,\ldots.
\end{equation}
$q_n$ is the Nome and is defined as
\begin{equation}
    q_n=\exp(-\pi K(\sqrt{1-\k^2_n})/K(\k_n)).
\end{equation}
Thus, $J_{nm}$ is given by (note that $J'=-J$ as the potential is
even):
\begin{equation}
\fl
J_{nm}=\frac{8\pi^3\wb}{K(\k_n)K(\k_m)}\sqrt{\frac{q_nq_m}{(1-2\k_n^2)(1-2\k_m^2)}}
    \sum_{\nu=0}^{\infty}\frac{(2\nu+1)^2(q_nq_m)^\nu}{(1+q_n^{2\nu+1})(1+q_m^{2\nu+1})},
\end{equation}
and, in consequence,
\begin{equation}\label{eq:hhard}
    \h_{nm}=\frac{\displaystyle\sum_{\nu=0}^{\infty}
    \frac{(2\nu+1)^2(q_nq_m)^\nu}{(1+q_n^{2\nu+1})(1+q_m^{2\nu+1})}}
    {\displaystyle\sqrt{\sum_{\nu=0}^{\infty}\left(\frac{(2\nu+1)q_n^\nu}{1+q_n^{2\nu+1}}\right)^2}
    \sqrt{\sum_{\nu=0}^{\infty}\left(\frac{(2\nu+1)q_m^\nu}{1+q_m^{2\nu+1}}\right)^2}}\,,
\end{equation}
\begin{equation}\label{eq:phard}
    \p_{nm}=\frac{\displaystyle\wm K(\k_n)}{\displaystyle\wn K(\k_m)}
    \sqrt{\frac{\displaystyle(1-2\k_n^2)\sum_{\nu=0}^{\infty}
    \left(\frac{(2\nu+1)q_m^\nu}{1+q_m^{2\nu+1}}\right)^2}
    {\displaystyle(1-2\k_m^2)\sum_{\nu=0}^{\infty}\left(\frac{(2\nu+1)q_n^\nu}{1+q_n^{2\nu+1}}\right)^2}}\,.
\end{equation}

\subsubsection{Soft $\phi^4$ potential. $s=-1$.}

Now, the orbit of an oscillator is given by:
\begin{equation}
 \fl   u_n(t)=\pm\wn\sqrt{\frac{2\k_n^2}{1+\k_n^2}}\cd\left(\frac{\wn t}{\sqrt{1+\k_n^2}},\k_n\right)=
    \pm\wn\sqrt{\frac{2\k_n^2}{1+\k_n^2}}\cd\left(\frac{2K(\k_n)}{\pi}\wt,\k_n\right),
\end{equation}
where $\cd$ is another elliptic function. Now, the breather
frequency $\wb$ and the modulus $\k_n$ are related through:
\begin{equation}
    \wb=\frac{\pi\wn}{2\sqrt{1+\k_n^2}K(\k_n)}.
\end{equation}
Following \cite{Abram}, the Fourier coefficients can be obtained:
\begin{equation}
    z_{2\nu+1,n}=\pm\frac{\pi}{K(\k_n)}\sqrt{\frac{2}{1+\k_n^2}}\,
    \frac{q_n^{\nu+1/2}}{1-q_n^{2\nu+1}},
\end{equation}
and, in consequence,
\begin{equation}
 \fl   J_{nm}=\frac{8\pi^3\wb}{K(\k_n)K(\k_m)}\sqrt{\frac{q_nq_m}{(1+\k_n^2)(1+\k_m^2)}}
    \sum_{\nu=0}^{\infty}\frac{(2\nu+1)^2(q_nq_m)^\nu}{(1-q_n^{2\nu+1})(1-q_m^{2\nu+1})},
\end{equation}
\begin{equation}\label{eq:hsoft}
    \h_{nm}=\frac{\displaystyle\sum_{\nu=0}^{\infty}
    \frac{(2\nu+1)^2(q_nq_m)^\nu}{(1-q_n^{2\nu+1})(1-q_m^{2\nu+1})}}
    {\displaystyle\sqrt{\sum_{\nu=0}^{\infty}\left(\frac{(2\nu+1)q_n^\nu}{1-q_n^{2\nu+1}}\right)^2}
    \sqrt{\sum_{\nu=0}^{\infty}\left(\frac{(2\nu+1)q_m^\nu}{1-q_m^{2\nu+1}}\right)^2}}\,,
\end{equation}
\begin{equation}\label{eq:psoft}
    \p_{nm}=\frac{\displaystyle\wm K(\k_n)}{\displaystyle\wn K(\k_m)}
    \sqrt{\frac{\displaystyle(1+\k_n^2)\sum_{\nu=0}^{\infty}
    \left(\frac{(2\nu+1)q_m^\nu}{1-q_m^{2\nu+1}}\right)^2}
    {\displaystyle(1+\k_m^2)\sum_{\nu=0}^{\infty}\left(\frac{(2\nu+1)q_n^\nu}{1-q_n^{2\nu+1}}\right)^2}}\,.
\end{equation}

\section{Demonstration of the necessity of an in-phase or
out-of-phase pattern for the stability of multibreathers}
\label{ap:b}

It was shown in Ref.~\cite{ACSA03} that a multibreather with
nearest neighbour coupling and whose vibration pattern is neither
 in-phase nor out-of-phase  is unstable independently
 of the hardness/sotftness of the on-site potential
and the sign of the coupling constant. However, this fact was not
demonstrated.

In this appendix, we will demonstrate the last assertion by making
use of a consequence of Sylvester's inertia law, called
Sylvester's Theorem \cite{Horn}. It establishes that, for a
positive definite matrix, the principal minors are positive, for a
negative definite one, the principal minors alternate their signs
when the dimension increases by one and, finally, a matrix is not
definite when none of the last conditions are fulfilled.

Let us suppose an homogeneous lattice. If the pattern of vector
$\s$ is different from the in-phase or staggered ones, there is
always a sequence $\{-1,1,1\}$ or $\{1,-1,-1\}$ on it. It is easy
to show that in both cases, the perturbation matrix has the form:
\begin{equation}
    Q=\left[\begin{array}{ccc} \ddots & & \\ & Q_0 & \\ & & \ddots
    \end{array}\right],
\end{equation}
with
\begin{equation}
    Q^0=\left[\begin{array}{ccc} -\g_0 & \g_0 & 0 \\ \g & 1-\g_0 & 0 \\ 0 & 1 &
    \bullet \end{array}\right],
\end{equation}
where $\bullet$ is a number that depends on the phase of the
particle adjacent to the pattern breaking sequence. As the
spectrum of $Q$ is invariant under rows/columns exchanges, the
block $Q^0$ can be placed at $Q_{1,1}$, and Sylvester's theorem
can be applied to $Q^0$. The 1st order minor is
$M^{(1)}=Q^0_{1,1}=-\g_0<0$, and the 2nd order one,
$M^{(2)}=Q^0_{1,1}Q^0_{2,2}-Q^0_{1,2}Q^0_{2,1}=-\g_0<0$. In
consequence, the matrix is not definite and there are positive and
negative eigenvalues.

Let us suppose now that an impurity is introduced in the pattern
breaking sites. The matrices for impurities in the first, second
and third sites are (note that we only consider the first
$(2\times2)$ submatrix as the result is independent on the third
row/column), respectively:
\begin{eqnarray}
 \fl   Q^0_1=\left[\begin{array}{ccc} -\g\p & \g \\ \g & -\g/\p+1
    \end{array}\right],\nonumber \\
    \lo{Q^0_2}=\left[\begin{array}{ccc} -\g/\p & \g \\ \g & (\h-\g)/\p
    \end{array}\right], \;
    Q^0_3=\left[\begin{array}{ccc} -\g_0 & \g_0 \\ \g_0 &
    \h/\p-\g_0 \end{array}\right],
\end{eqnarray}
and the corresponding minors are:
$\{M^{(1)}_1=-\g\p,M^{(2)}_1=-\g\p\}$,
$\{M^{(1)}_2=-\g/\p,M^{(2)}_2=-\g\h\}$,
$\{M^{(1)}_3=-\g_0,M^{(2)}_3=-\g_0\h/\p\}$. In consequence, the
matrices are not definite.

\section*{Acknowledgements}

This work has been supported by the MCYT/FEDER under the project
BMF2003-03015 / FISI. J Cuevas acknowledges an FPDI grant from `La
Junta de Andaluc\'{\i}a'.

\vspace{1cm}


\begin{thebibliography}{10}

\bibitem{MA94}
RS~MacKay and S~Aubry.
\newblock Proof of existence of breathers for time-reversible or
  \mbox{Hamiltonian} networks of weakly coupled oscillators.
\newblock {\em Nonlinearity}, 7:1623, 1994.

\bibitem{A97}
S~Aubry.
\newblock Breathers in nonlinear lattices: Existence, linear stability and
  quantization.
\newblock {\em Physica D}, 103:201, 1997.

\bibitem{MAF98}
JL~Mar\'{\i}n, S~Aubry, and LM~Flor\'{\i}a.
\newblock Intrinsic {L}ocalized {M}odes: {D}iscrete breathers. {E}xistence and
  linear stability.
\newblock {\em Physica D}, 113:283, 1998.

\bibitem{MS98}
RS~MacKay and JA~Sepulchre.
\newblock Stability of discrete breathers.
\newblock {\em Physica D}, 119:148, 1998.

\bibitem{MARIN}
JL~Mar\'{\i}n.
\newblock {\em {I}ntrinsic {L}ocalized {M}odes in nonlinear lattices}.
\newblock Ph{D} {T}hesis, University of {Z}aragoza (Spain), June 1997.

\bibitem{ACSA03}
JFR Archilla, J~Cuevas, B~S\'{a}nchez-Rey, and A~\'{A}lvarez.
\newblock Demonstration of the stability or instability of multibreathers at
  low coupling.
\newblock {\em Physica D}, 180:235, 2003.

\bibitem{CPAR02b}
J~Cuevas, F~Palmero, J~F~R Archilla, and F~R Romero.
\newblock Moving discrete breathers in a {K}lein--{G}ordon chain with an
  impurity.
\newblock {\em J. Phys. A: Math. and Gen.}, 35:10519, 2002.

\bibitem{Horn}
RA~Horn and CR~Johnson.
\newblock {\em Matrix Analysis}.
\newblock Cambridge University Press, 1985.

\bibitem{Sylvester}
JJ~Sylvester.
\newblock A demonstration of the theorem that every homogeneous quadratic
  polynomial is reducible by real orthogonal substitutions to the form of a sum
  of positive and negative squares.
\newblock {\em Phil. Mag.}, IV:138, 1852.

\bibitem{Tesis}
J~Cuevas.
\newblock {\em Localization and energy transfer in anharmonic inhomogeneus
  lattices}.
\newblock Ph{D} {D}issertation, Physics Faculty, University of Sevilla (Spain),
  2003.

\bibitem{AACR02}
A~\'{A}lvarez, JFR Archilla, J~Cuevas, and FR~Romero.
\newblock Dark breathers in {K}lein-{G}ordon lattices. {B}and analysis of their
  stability properties.
\newblock {\em New Jornal of Physics}, 4:72, 2002.

\bibitem{MJKA02}
AM~Morgante, M~Johannson, G~Kopidakis, and S~Aubry.
\newblock Standing waves instabilities in a chain of nonlinear coupled
  oscillators.
\newblock {\em Physica D}, 162:53, 2002.

\bibitem{Abram}
M~Abramowitz and IA~Stegun.
\newblock {\em Handbook of mathematical functions}.
\newblock Dover, New York, 1965.

\end{thebibliography}

\newcommand{\noopsort}[1]{} \newcommand{\printfirst}[2]{#1}
  \newcommand{\singleletter}[1]{#1} \newcommand{\switchargs}[2]{#2#1}

\end{document}